\documentclass[useAMS,usenatbib]{mn2e}
\usepackage{graphics}
\newcommand{\apj}{ApJ}
\newcommand{\apjl}{ApJL}
\newcommand{\mnras}{MNRAS}
\newcommand{\aj}{AJ}
\newcommand{\apjs}{ApJS}
\newcommand{\nat}{Nature}

\newcommand{\aap}{A\&A}
\newcommand{\aaps}{A\&AS}

\title[Star Formation at $z\ge 6.5$]{Constraints on Star Forming Galaxies at $z\ge 6.5$ from HAWK-I $Y$-band Imaging of GOODS-South}
\author[S. Hickey et al.]{Samantha Hickey$^{1,2}$\thanks{E-mail:
s.hickey@herts.ac.uk}, Andrew Bunker$^{3,2,4}$, Matt J.~Jarvis$^{1}$, Kuenley Chiu$^{2,4}$ \and and David Bonfield$^{1}$\\
$^{1}$Centre for Astrophysics, Science \& Technology Research Institute, University of Hertfordshire, Hatfield, Herts, AL10 9AB, UK\\
$^{2}$Anglo-Australian Observatory, Epping, NSW 1710, Australia \\
$^{3}$Astrophysics, Department of Physics, Keble Road, Oxford, OX1 3RH, UK \\
$^{4}$School of Physics, University of Exeter, Stocker Road, Exeter, EX4 4QL, 
U.K.}

\begin{document}

\date{Accepted ????. Received 2008 ???; in original form 2008 October xx}

\pagerange{\pageref{firstpage}--\pageref{lastpage}} \pubyear{2008}

\maketitle

\label{firstpage}

\begin{abstract}

  We present the results of our search for high-redshift Lyman-break
  galaxies over the GOODS-South field. We use {\em HST}-ACS data in
  $B$, $V$, $i'$ \& $z'$, VLT-ISAAC $J$ and $Ks$, {\em Spitzer}-IRAC
  3.6, 4.5, 5.8 and 8.0$\micron$ data in conjunction with the new
  HAWK-I $Y$-band science verification data to search for dropout
  galaxies in the redshift range $6<z<9$. We survey $\approx
  119$\,arcmin$^2$ to $Y_{AB}=25.7$ ($5\,\sigma$), of which
  $37.5$\,arcmin$^2$ reaches $Y_{AB}=25.9$. Candidate $z'$ and $Y$ drop-outs were selected
  on the basis of a colour cut of $(Y-J)_{AB}>0.75$\,mag and
  $(z'-Y)_{AB}>1.0$\,mag respectively. We find
  no robust Y-drops ($z\approx 9$) brighter than $J_{AB}<25.4$. In our
  search for $z'$-band dropouts ($z\approx6.5-7.5$), we identify four possible candidates, two with $z'$-drop colours and clear 
{\em Spitzer}-IRAC detections and two less likely candidates. We also identify two
  previously-known Galactic T-dwarf stellar contaminants with these 
colours, and two likely transient objects seen in the $Y$-band 
data. The
implications if all or none of our candidates are real on the
Ultra-Violet galaxy luminosity functions at $z>6.5$ are explored. We
find our number of $z'$-drop candidates to be insufficient based on the
expected number of $z'$ drops in a simple no-evolution scenario from
the $z=3$ Lyman-break galaxy luminosity function but we are 
consistent with the observed luminosity function at $z \approx 6$ (if
all our candidates are real). However, if one or both
of our best $z'$-drop candidates are not $z>6.5$ galaxies, this would
demand evolution of the luminosity function at early epochs, in the
sense that the number density of ultra-violet luminous star-forming galaxies at
$z>7$ is less than at $z\sim6$. We show that the future surveys to be conducted with the ESO VISTA telescope over the next five years will be able to measure the bulk of the luminosity function for both $z'$ and $Y$ drop-outs and thus provide the strongest constraints on the level of star-formation within the epoch of reionization.

\end{abstract}


\section{Introduction}
\label{sec:intro}

The identification of high-redshift galaxies is crucial to developing
our understanding of the early universe, galaxy evolution and the
epoch of reionisation.  One technique for selecting high-redshift
galaxies is to search for Lyman Break Galaxies (LBGs) through the
drop-out technique (Guhathakurta, Tyson \& Majewski 1990; Steidel, Pettini \& Hamilton 1995; Steidel et
al. 1996). This relies on finding a significant flux decrement between
two broadband filters indicating a spectral break short-ward of
Lyman-$\alpha$ at 1215.7\AA\ in the rest frame of the galaxy,
attributable to Lyman-$\alpha$ absorption from intervening clouds of
neutral hydrogen at lower redshifts (the Lyman-$\alpha$ forest). In
higher-redshift galaxies this ``Lyman break'' is pushed to longer
wavelengths.

Recent years have witnessed two key developments in our understanding
of the initial phases of galaxy formation. The first is the extension
of the Lyman-break selection technique to higher redshifts, so that
broad-band colours can be used to identify galaxies at $z=6$ (Stanway,
Bunker \& McMahon 2003; Bunker et al.\ 2004).  The Lyman-break
``$i'$-drop'' selection has been spectroscopically confirmed to find
galaxies at $z\approx 6$ (Bunker et al.\ 2003; Stanway et al.\
2007). The
second key development is the revelation that $z>6$ is a crucial
transition epoch in the history of the Universe. The discovery of
complete absorption by the Lyman-$\alpha$ forest (the Gunn-Peterson
trough) in the spectra of quasi-stellar objects (QSOs) at $z>6.2$ (Becker et al.\ 2001, Fan et al.\
2002) indicates a large neutral fraction of hydrogen, and the Wilkinson Microwave Anisotropy Probe (WMAP) results suggest that the Universe was entirely neutral at $z>10$
(Kogut et al.\ 2003) with a protracted period of reionization (Dunkley et al. 2009). Hence, exploring the epoch $6<z<10$ is crucial
if we are to understand what reionized the Universe, and thus set the
stage for galaxy formation at the end of the ``Dark
Ages''. Specifically, ultra-violet (UV) light from star forming galaxies during this
era has been proposed as the most likely reionization mechanism, as
the number density of high-redshift active galactic nuclei (AGN) appears too low to be solely
responsible (e.g. Dijkstra et al.\ 2004).


 Only a handful of galaxies are known at $z>6.5$
(Bouwens et al.\ 2005, 2008; Ota et al.\ 2008), selected through
Lyman-break broad-band imaging and also narrow-band imaging
for Lyman-$\alpha$ line emitting galaxies. The highest redshift spectroscopically confirmed galaxy currently known was discovered by Iye et al. (2006) at $z=6.96$ and the highest redshift object is the gamma-ray burst found at $z \approx 8.3$ (Tanvir et al. 2009).    The
frontier in spectroscopically confirmed galaxies and large robust
samples of Lyman Break galaxies is currently $z\approx 6$. The data
from the new VLT HAWK-I camera (Pirard et al., 2004; Casali et
al., 2006) in the $Y$-band filter in conjuction with the availability of ISAAC images over most of the ACS/WFC GOODS-South area
enables us to push this work even further in lookback time and to explore
the redshift range $6.5<z<9$. Constraints have also
been placed on the Ultra Violet luminosity function at $z\approx 7$ by
Mannucci et al.\ (2007).

\begin{figure}
\resizebox{0.48\textwidth}{!}{\includegraphics{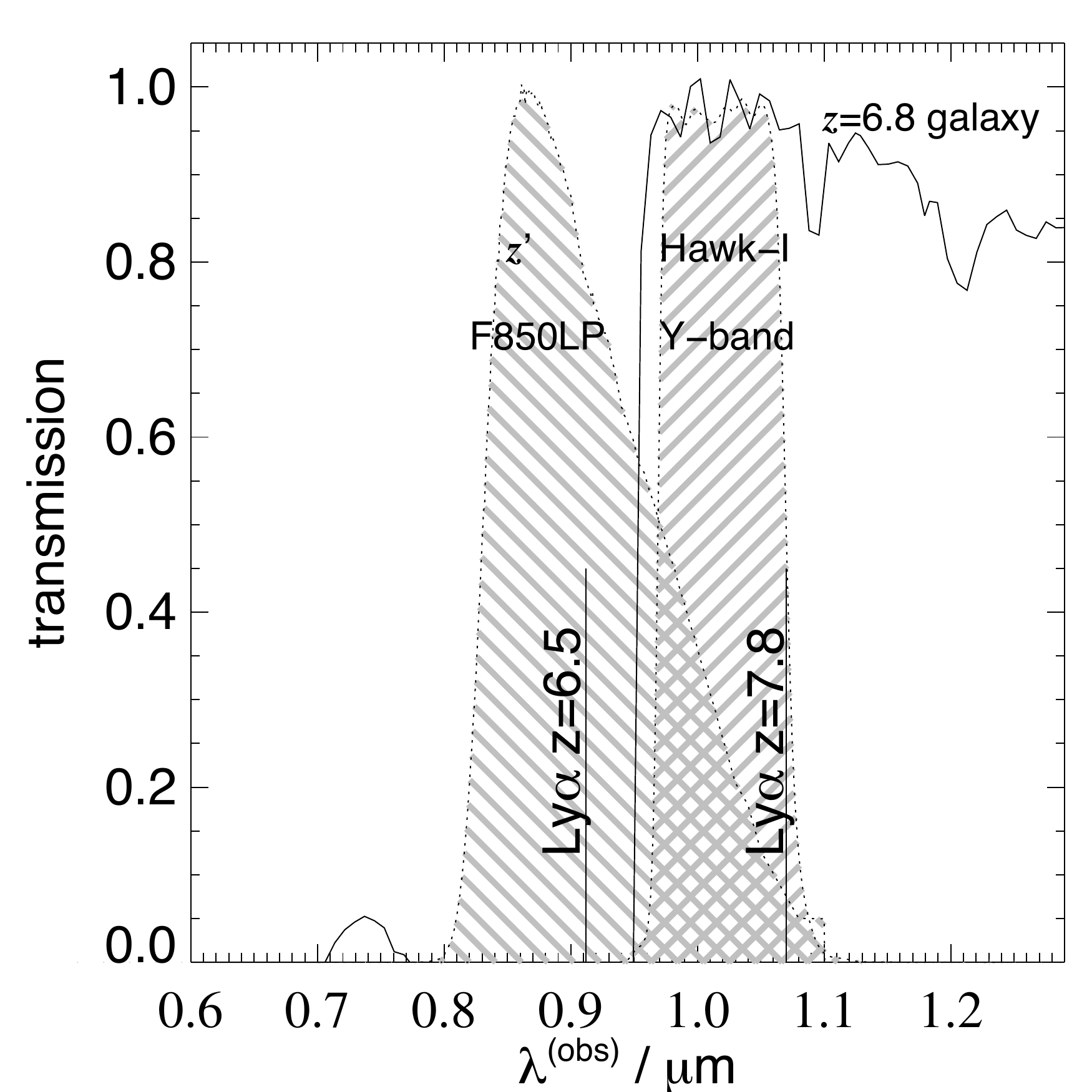}}
\caption{The ACS $z'-$ and HAWK-I $Y$-band passes over plotted on the
spectrum of a  $z=6.8$ galaxy with 100 Myr of constant
star formation and 0.2 Solar metallicity (solid line), illustrating the
utility of our two-filter technique for locating $z\approx 7$
sources.}
\label{fig:filters}
\end{figure}

To explore this population further we need a large sample of these
galaxies. This has been hampered until now by a lack of sensitivity in
the infra-red and the small fields of view available. Now with HAWK-I, an instrument with a large field of view  (7.5 x 7.5 arcmin$^2$) on the
VLT, an 8-metre class telescope and critically with the $Y$-band
filter, we can begin to increase the number of Lyman Break candidates
at $z>6.5$.  The $Y$-band filter is particularly important because it is a good discriminant of spectral breaks (a sharp cutoff) as opposed
to dust reddening, which appears as a more gradual slope, as it is centered on $1\mu$m which is close in
wavelength to the $z'$ filter's peak transmisison wavelength of
$\approx 0.9\mu$m (see Figure~\ref{fig:filters}). The $Y$-band filter is
also currently being used by UKIRT Infrared Deep Sky Survey (UKIDSS; Lawrence et al. 2007) to search for
QSOs at similar epochs (e.g. Venemans et al. 2007).

The structure of the paper is as follows. In Section~\ref{sec:obs} we
describe the imaging data. The construction of our catalogues,
survey completeness, our $z'$ and $Y$-drop selection and the likely
contaminants are discussed in Section~\ref{sec:selection}. In
Section~\ref{sec:discuss} we discuss our candidates and infer
constraints on the rest-frame UV luminosity
function of star-forming sources. Our conclusions are presented in
Section~\ref{sec:concs}. Throughout we adopt the standard
``concordance'' cosmology of $\Omega_M=0.3$, $\Omega_{\Lambda}=0.7$,
and use $h_{70}=H_0/70\,{\rm km\,s^{-1}\,Mpc^{-1}}$. We use the $AB$ magnitude system throughout (Oke \& Gunn 1983).

\section{Observations and Data Reduction}
\label{sec:obs}

We analysed $Y$-band observations from the ESO/VLT archive obtained as
part of HAWK-I science verification program 60.A-9284(B) (Fontana et
al.\ and Venemans et al.\ -- ``A deep infrared view on galaxies in the
early Universe'').  Two areas within the GOODS-South field were
imaged, with centres 03:32:41.0 $-$27:51:45 (pointing 1) and 03:32:29.6
$-$27:44:37 (pointing 2, both J2000). The seeing in $Y$-band was in the
range $0\farcs4-0\farcs7$ (FWHM) in the individual exposures, with the
stacked image having seeing of $\approx 0\farcs5$.

Individual exposure times were 30\,seconds, with 10 such exposures
averaged to form a single co-added frame of 300\,seconds.  A typical
sequence was 12 exposures (1\,hour), although we used sequences with
between 6 and 13 exposures. The telescope was dithered by 5-10$^{\prime\prime}$
between exposures. There were 18 sequences in pointing 1 (comprising
195 frames in all, a total of 16.25\,hours), and 14 towards pointing 2
(comprising 138 frames, amounting to 11.5\,hours).

The HAWK-I camera was read out in 4 quadrants, each of $2048\times
2048$ pixels with a scale $0\farcs106$\,pixel$^{-1}$, and initially
these quadrants were reduced individually before final mosaiking.
We subtracted the dark current and flat-fielded the data
using the average of $Y$-band twilight sky flats from the ESO HAWK-I
archive. For each sequence of $\sim$12 exposures, we then used the
{\tt DIMSUM} package within {\tt IRAF} to background-subtract the sky,
using the {\tt xmosaic} task. We measured the image shifts
interactively using a number of compact, bright but unsaturated
sources. Where there were 12 frames in a sequence, we used the average
of the 5 frames before and 5 after (with objects masked) to determine
the sky level. DIMSUM also rejected many of the cosmic rays present in
the images. However, there were various electronic read-out artifacts
visible which were pernicious and not removed by DIMSUM. Hence when we
combined the 14--18 final frames output by DIMSUM, we used
IRAF.imcombine, weighting by the exposure time, with the {\tt ccdclip}
rejection algorithm, using the gain and readnoise properties of the
detector and the Poisson noise of the sky background. A detector
artifact which was not effectively eliminated by this rejection was a
cross-talk effect, whereby ghost images appeared in the same detector
row as bright objects. As the apparent spatial position of the cross
talk artifacts on the sky remains fixed with respect to the objects
(i.e. it moves with the objects on the detector during the dithering
pattern), this was not rejected in the co-addition. We eliminated this
later by visual inspection and by masking detector rows affected by
bright objects.

For each quadrant and for both pointings we then determined the astrometry
through comparison with the ESO ISAAC $J$-band image (Retzlaff et al.\
{\em in prep.}) \footnote{We used version 2 -- available from
  http://archive.eso.org/archive/adp/GOODS/ISAAC$\_$imaging$\_$v2.0}
using 60--110 objects in each quadrant. A quadratic fit to the
distortion produced residuals of $\sim 0\farcs1$ between the
coordinates in the $J$- and $Y$-band images. The {\tt iraf.wregister}
routine was used to linearly interpolate the $Y$-band images to map
on to the same pixel grid as the $J$-band, with a pixel scale of
$0\farcs 15$.

We summed the exposure maps output by DIMSUM using the same measured
shifts, creating a map of the total exposure time as a function of
position on the sky, and correcting the astrometric distortions as
described above. This was then used to inverse-variance weight the
images when we combined the four quadrants in both pointings to form a
final image mosaic.  The 119\,arcmin$^2$ of the
$Y$-band image covered most of the ESO $J$-band and {\em HST} ACS
GOODS images (97.5\% of the HAWK-I $Y$-band image overlapped with the
ESO $J$-band). We used the GOODS team reductions of the ACS images
(Giavalisco et al.\ 2004), consisting of F450W $B$-band, F606W
$V$-band, F775W $i'$-band and F850LP $z'$-band. The GOODS images had
been drizzled from the original ACS pixel scale of $0\farcs 05$ on to a
grid of $0\farcs 03$ pixels. We used version 2.0 of the ACS GOODS
images\footnote{The GOODS ACS v2 images are available at {\tt
    http://archive.stsci.edu/pub/hlsp/goods/v2/}} and adopted the AB
magnitude zeropoints re-determined for the v2 release.

At the time of writing, the formal $Y$-band zero point for HAWK-I
was not available, so we determined the photometric zero point of the
$Y$-band by measuring the $Y-J$ colours of objects in identical
apertures of 1\,arcsec diameter, and setting the average AB colour of sources with flat spectra between the $z'$ and $J$-band to
be zero. Using our computed zero point, the AB magnitude in the
$Y$-band is given by
\[
Y_{AB} = 30.46-{\rm (ap.\ corr.)}-2.5\log_{10}({\rm count\ rate})\,{\rm mag}
\]
where ``ap\ corr'' is the aperture correction in magnitudes, and
``count rate'' is the number of counts per second recorded.  We
determined the aperture correction using a 1\,arcsec-diameter aperture
to be $0.4$\,mag for compact but unsaturated sources (see Section
\ref{sec:catalog}).

The $5\,\sigma$ detection limit for a
compact source in a 1\,arcsec-diameter aperture is $Y_{AB}=25.7$\,mag,
however pointing 1 is slightly deeper and reaches
$Y_{AB}=25.9$\,mag ($5\sigma$).  In all, we survey $37.5\,{\rm arcmin}^{2}$ to a
maximum $5\,\sigma$ depth of $Y_{AB}=25.9$, and an area of $90.6\,{\rm
  arcmin}^{2}$ to $Y_{AB}<25.7$. The total area surveyed to
$Y_{AB}<25.5$ was $115.6\,{\rm arcmin}^{2}$.

In order to identify $Y$-band and $z'$-band drop-outs through their
extreme $(Y-J)$ or $(z'-Y)$ colours, we compare the
HAWK-I $Y$-band image with the GOODS team $J$-band and $z'$-band
images (taken with VLT-ISAAC and {\em HST}-ACS respectively). The
$5\,\sigma$ limiting magnitudes in GOODS ACSv2 measured in $1\farcs0$ diameter apertures are $B_{AB}=27.2$,
$V_{AB}=26.90$, $i'_{AB}=26.09$ and $z'_{AB}=26.14$. We block-averaged
the drizzled $0\farcs03$ ACS pixels $5\times 5$ to produce a $z'$-band
frame which was registered to the VLT/ISAAC $J$-band. The limiting
magnitudes in $1\farcs0$ apertures for the VLT-ISAAC images are
$J=25.2$ and $K_S=24.7$ ($5\,\sigma$ AB magnitudes). We also made use
of imaging of the GOODS-South field obtained with the Infrared Array
Camera (IRAC; Fazio et al. 2004) on board the {\em Spitzer Space Telescope} which was
conducted as part of the GOODS Legacy program (PID 194, PI Dickinson). IRAC uses four broad-band filters with central
wavelengths at approximately 3.6$\mu$m, 4.5$\mu$m, 5.8$\mu$m and 8.0$\mu$m (channels 1--4). The data were taken over two observing epochs,
with the telescope roll angle differing by $180^{\circ}$, and we used
the v2 and v3 reductions of the GOODS team, with the data drizzled
onto a $0\farcs6$ grid from the original $1\farcs2$ pixels.  We measure 5$\sigma$ limiting magnitudes of  24.76, and 24.87 for IRAC channels 1 and 2 respectively, measured in $2\farcs4$
diameter apertures, and 22.77 and 22.81  for IRAC channels 3 \& 4 measured
in $3\farcs 0$ \& $3\farcs7$ diameter apertures respectively (these
limits include aperture corrections of $\approx 0.7$\,mag appropriate
for unresolved sources, e.g. Eyles et al. 2005; 2007).

\section{Selection of $z'$-drop and $Y$-drop Candidates}
\label{sec:selection}

\subsection{Construction of Catalogues}
\label{sec:catalog}

Candidate selection for all objects in the field was performed using
version 2.4.6 of the SExtractor photometry package (Bertin \& Arnouts
1996). As we are searching specifically for objects which drop-out at
short wavelengths through the Lyman-$\alpha$ forest absorption, we used
SExtractor in dual-image mode, detecting objects in the longer-wavelength band
and measuring the photometry within the same spatial apertures in the
drop-out band(s).  We produced separate catalogues for the $Y$-band drop
outs (using the $J$-band as the detection image) and the $z'$-band
drop-outs (using the $Y$-band as the detection image). To reduce the
number of spurious sources in the noisy edge regions (where few frames
overlap) we used the exposure maps as input weight maps for
SExtractor.  For object identification, we adopted a limit of at least
5 contiguous pixels above a threshold of $2\sigma$ per pixel (on the
data drizzled to a scale of 0\farcs15~pixel$^{-1}$). Spurious detections close to the
noise limit were later eliminated through colour cuts and visual inspection. 

As high redshift galaxies in the rest-UV are known to be compact (e.g., Ferguson et
al.\ 2004; Bouwens et al.\ 2004), we used fixed
circular apertures $1\farcs0$ in diameter to select our candidates and
corrected the aperture magnitudes to approximate total magnitudes for
each filter through an aperture correction, determined from bright
compact sources. These were measured to be 0.07\,mag in $z'$-band, 0.4\,mag in $Y$-band and
0.4\,mag in $J$-band. For compact sources, this approach is simpler
and more reproducible than using the SExtractor curve-of-growth total
magnitudes.

Section 9 in the version 2.0 of the ISAAC $J$-band image produced a
number of spurious sources when running SExtractor in dual image mode
and weighting on the exposure map. This appeared to be due to an
overconfidence in the weight map on section 9. This was corrected for
by weighting on the background for this section alone and using the
weight map for the rest of the $J$-band image.

\subsection{Completeness}
\label{sec:comp}

The completeness corrections for both the $Y-$ and the $J$-band images
were measured in the following way. Approximately 5000 artificial
compact sources were created with diameters of 3 pixels and spanning
magnitudes between 20 and 30.  These objects were then convolved with
the PSF and added in to the original image. The new images were run
through SExtractor again, using the same criteria that was employed to
generate our object list. The resulting catalogues were compared with
the list of input {\em 'fake'} sources and a detection was considered
to be made if a source was found within 5 pixels of its input position
and had a magnitude correct to within a factor of 2 of the input
flux. The number of detected sources was then compared to the number of
input sources for each magnitude bin.

\begin{figure}
\resizebox{0.48\textwidth}{!}{\includegraphics{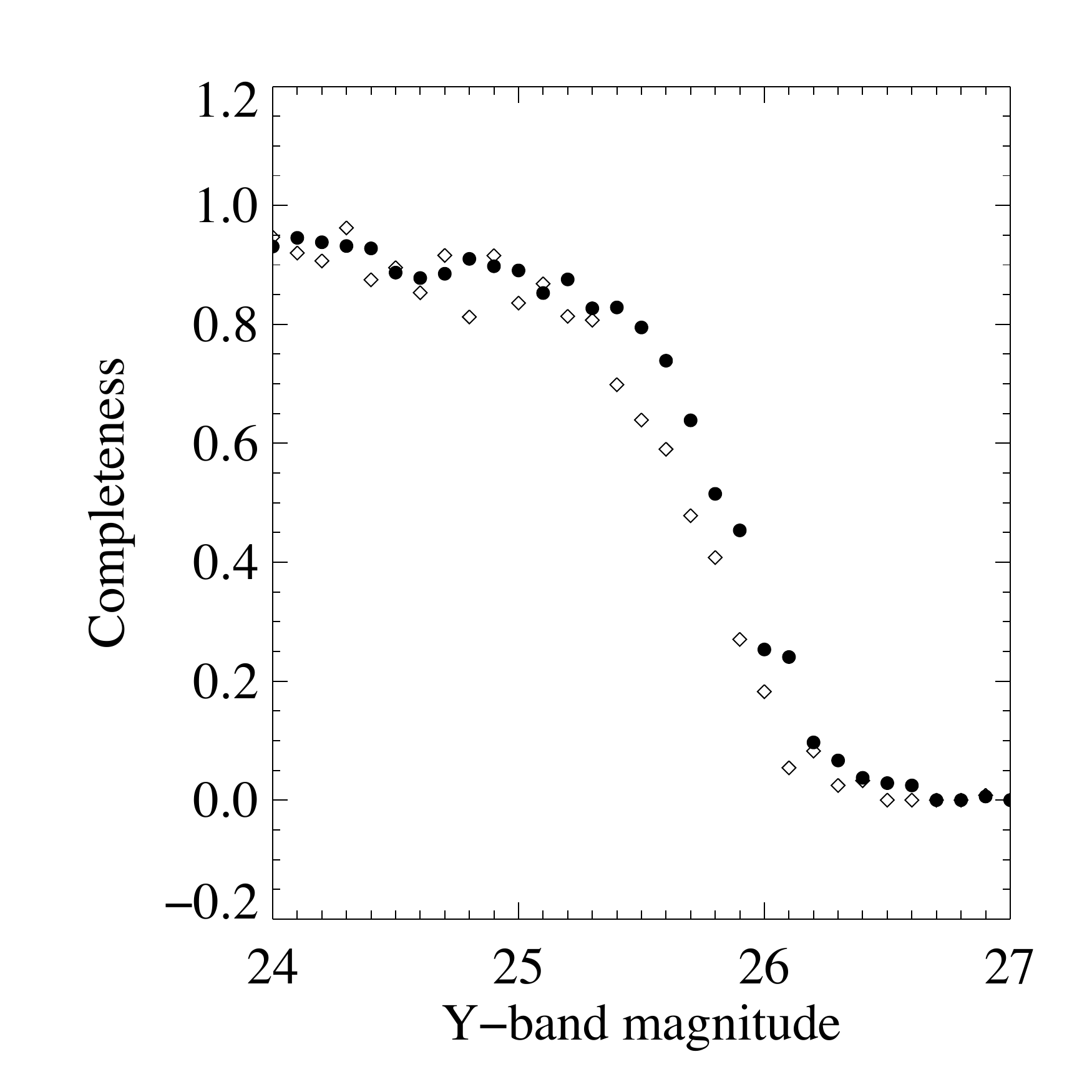}}
\caption{The $Y$-band completeness with SExtractor parameters of at least 5 pixels with $S/N > 2\sigma$. Pointing 1 is denoted by the filled circles and Pointing 2 by the open diamonds.}
\label{fig:Ycomp}
\end{figure}

\begin{figure}
\resizebox{0.48\textwidth}{!}{\includegraphics{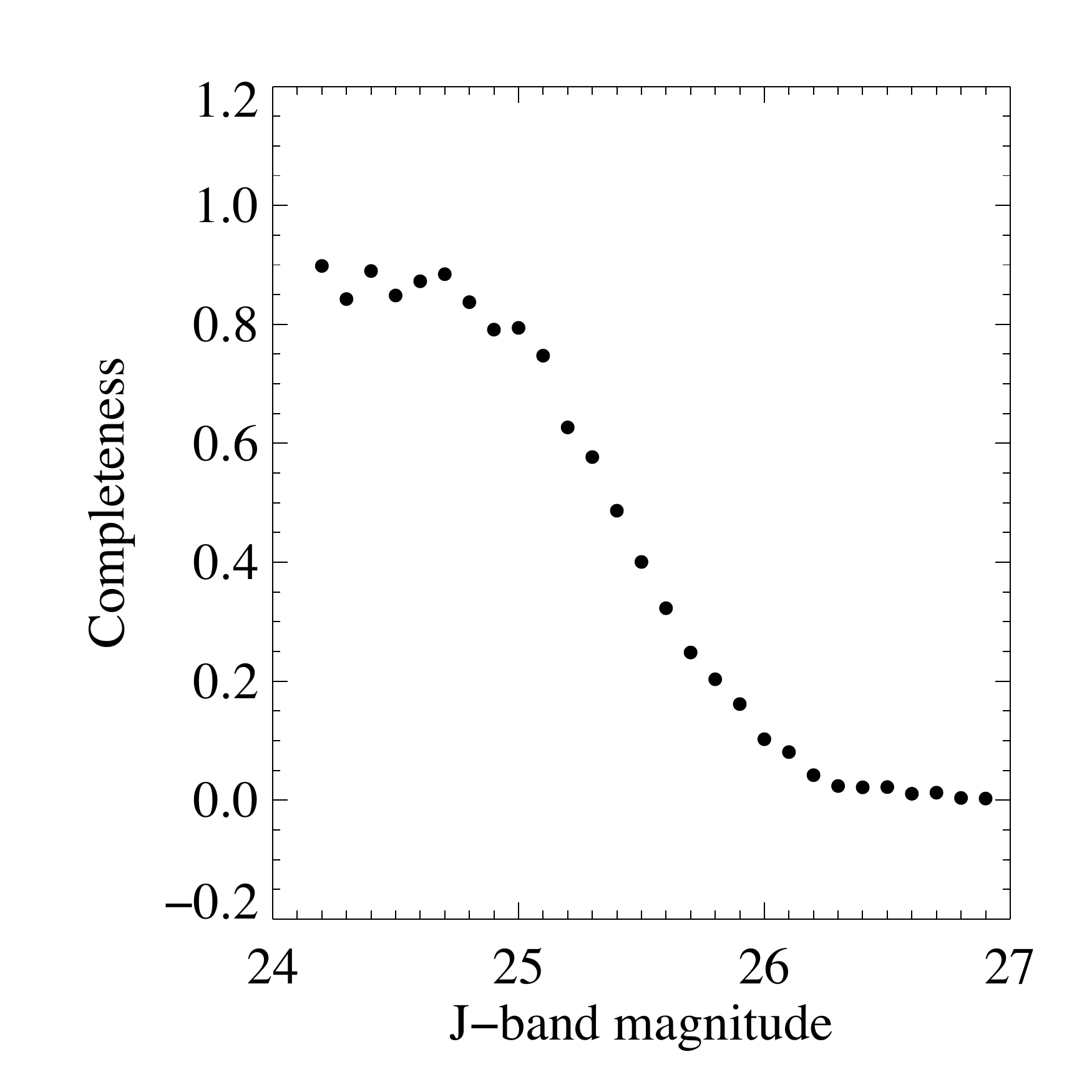}}
\caption{The $J$-band completeness with SExtractor parameters of at least 5 pixels with $S/N > 2\sigma$}

\label{fig:Jcomp_5}
\end{figure}

As described earlier, the $Y$-band image consisted of two individual
pointings of unequal depth, to measure an accurate completeness limit,
the calculations were determined for both pointings separately.
The filled circles in Figure \ref{fig:Ycomp} show the $Y$-band image for
Pointing 1 is $\sim 95$ per cent complete down to a magnitude of
$Y_{AB}=24.0$ and is 50 per cent complete at a magnitude of
$Y_{AB}=25.9$ over the deepest area. The $Y$-band image for Pointing 2 (denoted by open diamonds in figure~\ref{fig:Ycomp}) is $\sim 95$ per
cent complete down to a magnitude of $Y_{AB}=24.0$ and is 50 per cent complete
at a magnitude of $Y_{AB}=25.7$ over the deepest area in that pointing. We also estimate the completeness of the $J-$band image with a similar method. Figure
\ref{fig:Jcomp_5} shows the $J$-band image is $\sim 90$ per cent complete to
$J_{AB}=24$ and is 50 per cent complete at $J_{AB}=25.4$.

We calculated our $5\,\sigma$ limit in the $Y$-band to be 25.9 in pointing 1,
which agrees with our 50 per cent completeness limit (Fig.~\ref{fig:Ycomp}).  We tested the SExtractor
parameters by reducing our detection thresholds and found they had
little or no effect on our completeness implying that our original
parameters of at least 5 pixels at $2\,\sigma$ or above were
reasonable and did not eliminate credible sources from
our selection.

\begin{figure}
\resizebox{0.48\textwidth}{!}{\includegraphics{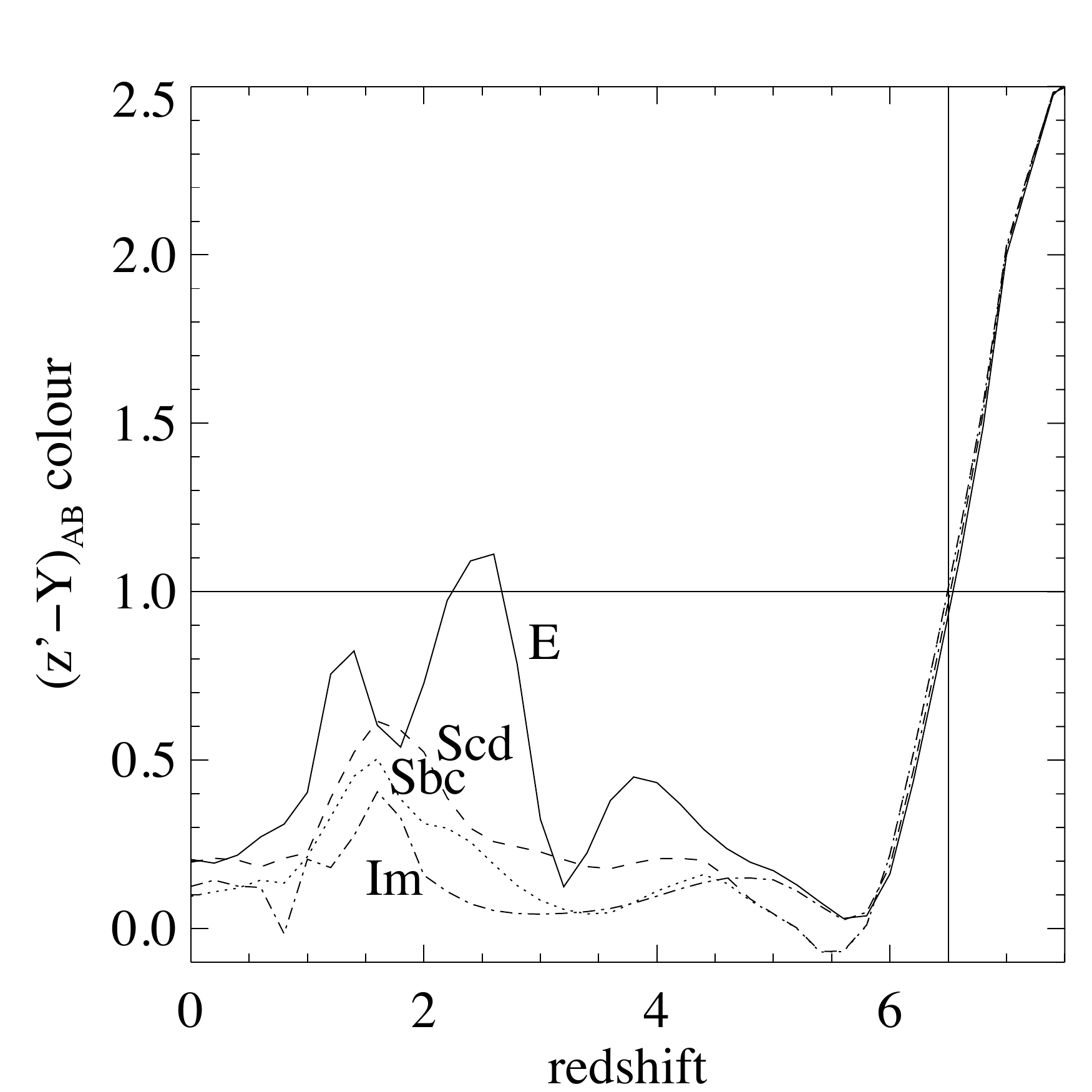}}
\caption{Model colour-redshift tracks for galaxies with non-evolving
stellar populations (from Coleman, Wu \& Weedman 1980 template
spectra). The contaminating `hump' in the $(z'-Y)$ colour at
$z\approx 2$ arises when the Balmer break and/or the 4000\,\AA\
break redshifts beyond the $i'$-filter. }
\label{fig:tracks}
\end{figure}

\subsection{$z$-drop Candidate Selection}

\begin{figure}
\resizebox{0.48\textwidth}{!}{\includegraphics{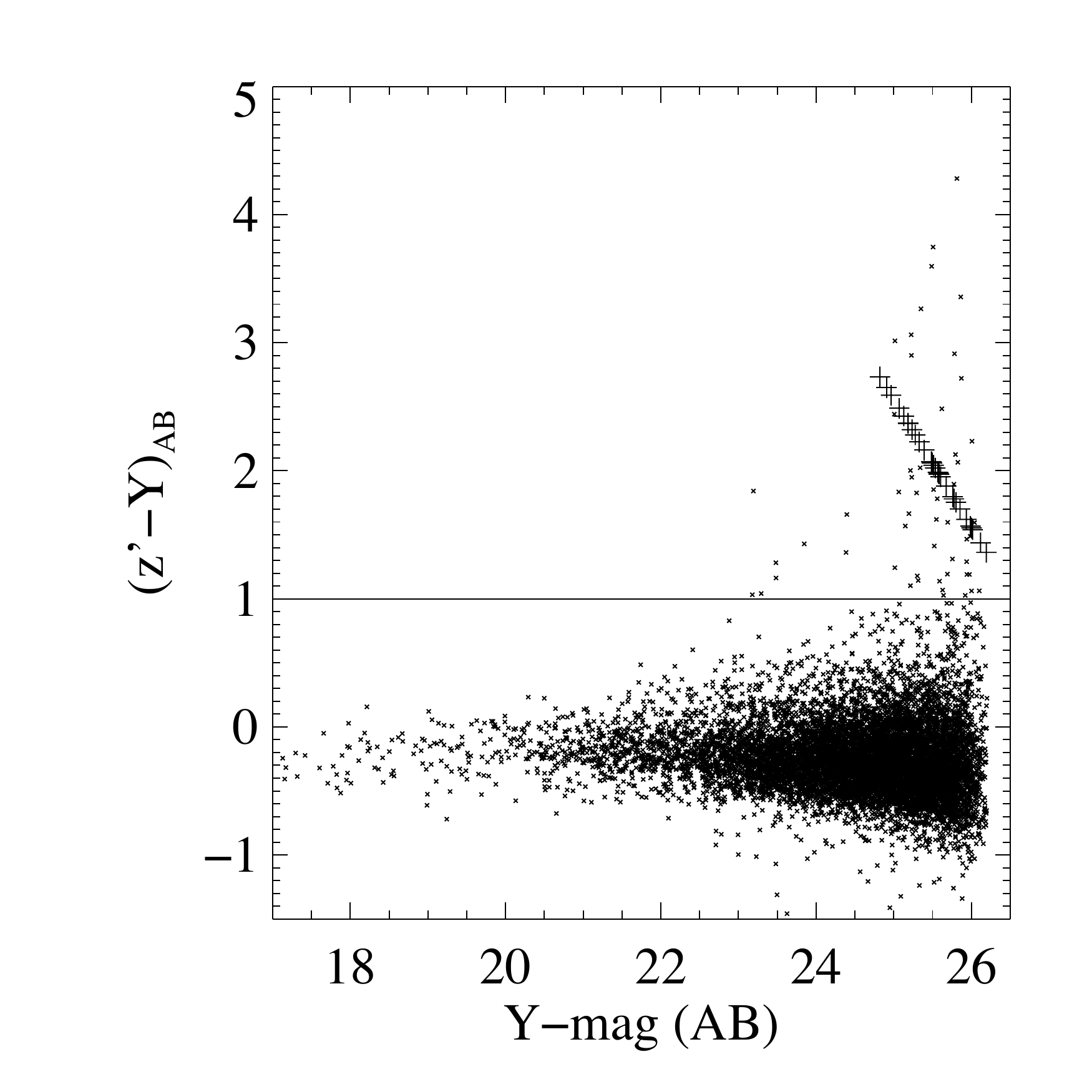}}
\caption{$z'-Y$ colours of all objects in the HAWK-I data. The  plus signs are 2\,$\sigma$ lower limits
on the $(z'-Y)$ colour where objects are undetected in
the $z'$-band.}
\label{fig:colmagyz}
\end{figure}

\begin{table*}
\caption{ Properties of our 4 $z'$-drop candidates, magnitudes are listed with an aperture correction applied as described in the text. Where the candidate is undetected we have quoted the $2\sigma$ limiting magnitude. Object 2200 is probably a lower redshift contaminant at $3.25<z<3.85$ as discussed in the text.  Objects 9136 and 9697 are our most convincing $z'$-drop candidates.}
\begin{tabular}{c|c|c|c|c|c|c|c|c|c|c}
\hline\hline

Our ID & 2200 & 9136 & 9266 & 9697\\
\hline
\hline
RA \& Dec &  03 32 25.3  -27 52 30.7 & 03 32 17.4  -27 43 43.0 & 03 32 19.2  -27 43 33.4 & 03 32 22.7  -27 43 00.8\\
$z'_{AB}$  & 26.87 $\pm$ 0.31 & 27.19 $\pm$ 0.38  & $>$28.1 & 27.84 $\pm$ 0.65\\
$Y_{AB}$ & 25.70 $\pm$ 0.14 & 25.90 $\pm$ 0.18  & 25.94 $\pm$ 0.19 & 25.29 $\pm$ 0.10\\
$J_{AB}$ & 25.23 $\pm$ 0.16 & 24.98 $\pm$ 0.23 & $>$26.18 & 26.0 $\pm$ 0.42\\
$K_{AB}$ & 23.8 $\pm$ 0.07 & 24.98 $\pm$ 0.32 & $>$ 25.42 & $>$25.42\\
$3.6_{AB}$ & 21.4 $\pm$ 0.01 & 23.6 $\pm$ 0.054 & $>$25.76 & 24.01 $\pm$ 0.07\\ 
$4.5_{AB}$ & 21.1 $\pm$ 0.01 & 23.55 $\pm$ 0.07 & $>$25.87 & 24.35 $\pm$ 0.13\\
$5.8_{AB}$ & 20.53 $\pm$ 0.02 & $>$23.77 & $>$23.77 & $>23.77$\\
$8.0_{AB}$ & 20.46 $\pm$ 0.02 & 22.24 $\pm$ 0.28 & $>$23.81 & $>$23.81\\
$(z'-Y)_{AB}$ & 1.17 & 1.29 & 2.10 & $>$2.47\\

\hline

\end{tabular}
\label{tab:zcandidates}

\end{table*}

For our $z'$-drop selection, our SExtractor catalogues revealed 278 candidates with aperture corrected colours $(z'-Y)_{AB}>1$\,mag and $S/N>5$ in the $Y$ band. 
We expect that many of these candidates will be low-redshift
interlopers such as low-mass stars or red galaxies at $z\sim 2$, which
can produce large $(z'-Y)$ colours (in particular due to 4000\,\AA\
and Balmer breaks -- see Figure~\ref{fig:tracks}). We chose this colour cut of $(z'-Y)_{AB}>1$\,mag because although it does not eliminate all of the low redshift interlopers it does omit a significant fraction without excluding potential candidates at $z>6.5$. We show the distribution of $z'-Y$ colours for all of our detected objects in Figure~\ref{fig:colmagyz}.
In order to eliminate
obvious low-redshift contaminants, we compared our list of
$z'$-drop candidates to the GOODS MUSIC catalogue (Grazian et al.\
2006) with a matching radius of $0\farcs36$. This catalogue provides photometry from HST-ACS, {\em Spitzer}-IRAC, and ground-based $U$-band and VLT-ISAAC $JHK_{s}$
imaging, with PSF-matching used to determine accurate colours. The
GOODS-MUSIC catalogue includes photometric redshift estimates derived
from the 14-band photometry, and the catalogue is a combination of a
$K_{s}$-band and $z'$-band selection. 

Of the 278 candidates, 101 appeared in the GOODS-MUSIC catalogue, mostly
with photometric redshifts $z_{phot}=1-2.5$, although there were two
with $z_{phot}=6.9$ which were identified by Mannucci et al. (2007) as brown dwarfs -- these objects are further discussed in Section~\ref{sec:otherwork}.

There was also one other candidate with $z_{phot}>5$ that was identified in our original
search but was found to have a match to GOODS-MUSIC catalogue object
30046. It has detections in the $i'$, $z'$, $Y$, $J$,
$Ks$ and {\em Spitzer} bands with very strong emission at the {\em
Spitzer} wavelengths. Its photometric redshift of $z_{phot}=5.14$,
along with its detection in the $i'$-band and its non-detection at the
shorter ACS wavelengths, leads us to classify this object as a
possible $V$-drop and therefore remove it from our sample of $z'$-drop
candidates.

\begin{figure}
\resizebox{0.48\textwidth}{!}{\includegraphics{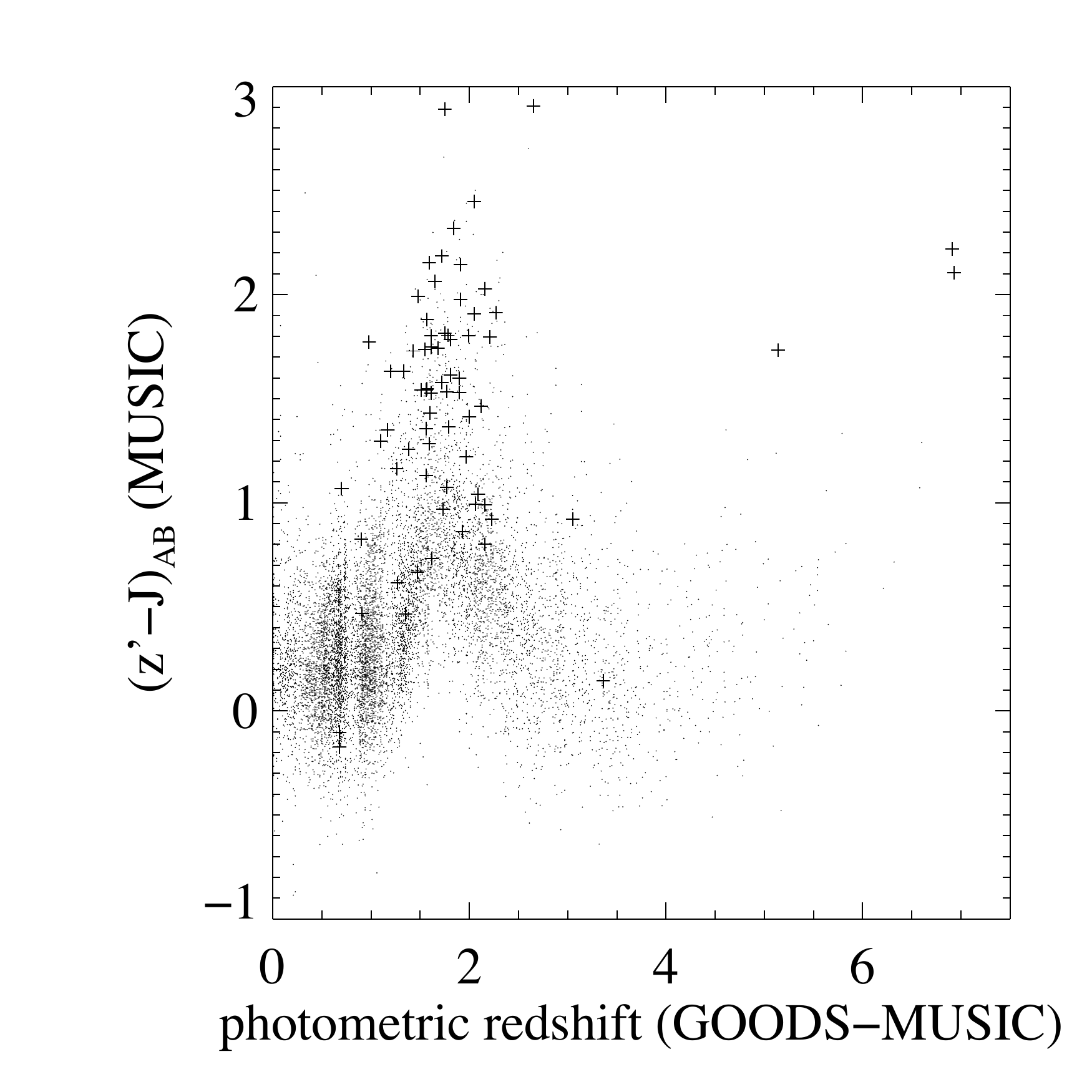}}
\caption{$z'-J$ colour versus the photometric redshift derived from the GOODS-MUSIC catalogue. The $z'$-drop out candidates that satisfy our colour selection criteria are marked with crosses. All of the sources in the GOODS-MUSIC catalogue with $S/N>3$ in the $J$-band are denoted by dots}
\label{fig:GOODS_MUSIC}
\end{figure}

We confirmed that all of the $z_{phot}<5$ matches to GOODS-MUSIC had
detections in one or more of the deep ACS $B$-, $V$- and $i'$-bands,
ruling out high-redshift interpretations due to the absence of a break
at Lyman-$\alpha$. In Figure~\ref{fig:GOODS_MUSIC} we show the $(z'-J)$
colours of our $z'$-drop candidates with GOODS-MUSIC matches
over-plotted with the full GOODS-MUSIC catalogue (with a $J$-band
threshold of $S/N>3$). As can be seen, most of our sources have
$(z'-J)_{\rm MUSIC}>1.0$, as would be expected from our selection of
$(z'-Y)>1.0$, and the bulk lie at $z_{phot}\approx 2$ (as would be
expected for the interloper populations -- see
Figure~\ref{fig:tracks}).
The $z'-$drop candidates that do not lie at $z \sim 2$ or $z>6$ are largely attributable
to the GOODS-MUSIC catalogue dealing with total magnitudes, whereas we
chose to use aperture magnitudes (more accurate for the expected
compact nature of high-redshift galaxies), with large low-redshift
galaxies having a greater aperture correction than we adopted. Also,
colour gradients within galaxies mean that aperture photometry may
select red regions of galaxies (e.g. spiral bulges) as $z'$-drop
candidates; the HST-ACS $z'$-band has better resolution than the
ground-based $Y$-band so we are sometimes susceptible to edges of
large objects as spurious candidates.

For the 177 $z'$-drop candidates which did not lie within $0\farcs36$ of a GOODS-MUSIC source, we visually inspected all four HST-ACS wavebands, as well as the HAWK-I $Y$-band and the ISAAC $J$ and $K_{s}$ bands, to ascertain whether the $Y$-band detection was real,
and if there was any detection at other wavelengths. Flux in
the ACS $B$-, $V$- or $i'$-bands would be incompatible
with the source being a $z'$-drop Lyman-break galaxy at $z>6.5$.  We found that 17.5 per cent of the remaining 177 candidates were detector artifacts (most frequently the cross-talk effect due to a bright object in the same detector row, manifesting a positive-negative dipole signal). Ghost image haloes around bright stars accounted for another 14 per cent, and 14 per cent again of the candidates were unreliable due to falling in regions of excess noise (despite using exposure weight maps to cut-down on spurious detections). 

To verify the reality of our six remaining $z'$-drop candidates, we split the $Y$
band data in to two halves (in time) and combined the first half and
the second half of the data separately. We then measured the $Y$ band
magnitudes of the candidates in both halves of the data (along with
some reference star (in case the seeing or magnitude zeropoints
differed over time) to check for consistency. Two of the six remaining
candidates were eliminated during this process as they were only
visible in the second half of the data.  
These objects appeared very bright
in the $Y$-band and were undetected in all of the other bands. This
prompted us to examine the individual images which
revealed the two sources to be time variable. Each night's data was
combined separately and the photometry on the individual nights were
measured for each candidate. Both objects were undetected on
2007-10-17, 2007-10-18, and 2007-10-19 but were visible on 2007-12-01
and 2007-12-02 with magnitudes from $Y_{AB}=24.77-24.88$. The two transient objects have coordinates of $\alpha = 03^{h}32^{m}54.4^{s}$  $\delta=-27^{d}53^{m}35.7^{s}$ and $\alpha=03^{h}32^{m}33.5^{s}$ $\delta=-27^{d}49^{m}38.3^{s}$.

Hence we find that 46.7 per cent of the 177 $z'$-drop candidates
without GOODS-MUSIC matches are spurious. A comparable fraction (50
per cent) had detections visible in the HST-ACS images; it is probable
that they did not have corresponding GOODS-MUSIC matches because of
the small matching radius we adopted ($0\farcs36$) to cut down on
multiple matches. Small astrometric shifts in some regions, the
$z'$-band and $K$-band magnitude cuts in GOODS-MUSIC, and spatially
extended galaxies with colour gradients account for these objects not
having GOODS-MUSIC matches.  Hence, from
an initial colour selection resulting in 278 objects, we have 4
remaining $z'-$drop-out candidates without GOODS-MUSIC matches.
Table \ref{tab:zcandidates}
contains our list of $z'$-drop candidates.

 Some of our $z'$-drop candidates may simply meet our selection criteria due to photometric scatter. To assess how significant this may be in our dataset we use our parent catalogue of $Y-$band detected sources over the 90.6 arcmin$^{2}$, which reaches a 5$\sigma$ depth of $Y_{AB} = 25.7$~mag (see section~\ref{sec:obs}), and randomly redistributed the magnitudes according their corresponding uncertainties in both the $z'$ and $Y-$bands and calculated the new $z'-Y$ colour for each source. We find, on average, that we detect nine objects purely due to photometric scatter in the parent catalogue. However, if we then take the fraction of the total sources that we exclude through our subsequent cuts through matching to the HST-ACS images etc., then we expect approximately 0.2 sources to be spurious in our final candidate list. Therefore, it is conceivable that one of our sources is a product of photometric scatter. Due to the lack of detection in any band other than the $Y$-band then Object 9266 is plausibly spurious.

\subsection{$Y$-drop Candidate Selection}

\begin{table*}
\caption{ Properties of our 4 $Y$-drop candidates, magnitudes are listed with an aperture correction applied as described in the text. Where the candidates are undetected in $Y$ we have quoted our $2\sigma$ limiting magnitude}

\begin{tabular}{c|c|c|c|c|c}
\hline\hline

Our ID & 2058 & 4551 & 5512 & 4532 \\
\hline
\hline
RA \& Dec & 03 32 27.6  -27 51 04.1  & 03 32 16.2  -27 47 39.1  & 03 32 27.5  -27 46 14.7 & 03 32 48.3  -27 47 39.9\\
$Y_{AB}$ & $>$26.6 & $>$26.6 & $>$26.6 & 27.47 $\pm$ 0.6438 \\
$J_{AB}$ & 25.37 $\pm$ 0.19 & 25.07 $\pm$ 0.19 & 25.05 $\pm$ 0.19 & 25.37 $\pm$ 0.19\\
$K_{AB}$ & $>$25.42 & $>$25.42 & $>$25.42 & $>$25.42\\ 
$3.6_{AB}$ & $>$25.76 & $>$25.76 & $>$25.76 & $>$25.76\\ 
$4.5_{AB}$ & $>$25.87 & $>$25.87 & $>$25.87 & $>$25.87\\
$5.8_{AB}$ & $>$23.77 & $>$23.77 & $>$23.77 & $>23.77$\\
$8.0_{AB}$ & $>$23.81 & $>$23.81 & $>$23.81 & $>$23.81\\
$(Y-J)_{AB}$ & $>$1.23 & $>$1.53 &  $>$1.55 & 2.1 \\

\hline

\end{tabular}
\label{tab:ycandidates}
\end{table*}

The $Y$-drop candidate selection was carried out in the same manner as
the $z'$-drop selection.  Our criteria for the
$Y$-drop candidates were a colour difference of $(Y-J)_{AB}>0.75$, a
signal-to-noise ratio $S/N>5$ in the $J$-band and a value in both the $Y$- and
$J$-band exposure maps equivalent to a minimum of 2.5\,hours of
observation in the $Y$ band. This selection yielded a list of 133
possible $Y$-drop candidates. We then compared our list to the GOODS
MUSIC catalogue to eliminate the low redshift interlopers from our
selection. In all, 98 of our 133 objects had matches within
$0\farcs36$ of our candidates. One of these objects had no
detection in the bands $B$, $V$ or $i'$ so we kept it in our candidate
list. This resulted in 37 $Y$-drop candidates. These remaining objects
were inspected more closely with postage stamps in $B$, $V$, $i'$,
$z'$, $Y$, $J$ and $K_{s}$ bands. We found 16 per cent of our
remaining candidates to be ghost image haloes around bright stars, 27
per cent were detections picking up on the edges of bright
galaxies in the $J$-band. Another 3 per cent fell on noisy regions
of the $Y$-band and 43 per cent of our candidates had visible ACS
detections. From our original list of 133 candidates only 4 possible
$Y$-drops remain, and these are listed in Table \ref{tab:ycandidates}.

\section{Discussion}
\label{sec:discuss}

\subsection{Candidates}

We have four $z'$-drop candidates remaining after eliminating artifacts
and low-redshift interlopers. Our candidates span a range
$Y_{AB}=25.3-26.0$ (after applying our aperture correction),
two of which have $>2\,\sigma$ detections in the $z'$-band. Three of the
candidates have strong detections in the IRAC wavebands. We obtained
photometric redshifts for these objects because the detection at the
IRAC wavelengths increases the likelihood of a valid detection and
also improves the accuracy of the photometric redshift solutions. We
used the publicly available software {\sc Hyperz} \footnote{ Hyperz is
  available at http://webast.ast.obs-mip.fr/hyperz/} (Bolzonella, Miralles \& Pell{\'o} et al. 2000) to derive our photometric redshift estimates in the redshift
range $z=0-9$. We allowed visual extinction values between $A_V=0-4$
and assumed the Calzetti (1997) reddening law. We used the eight Bruzual
\& Charlot (2003) template spectra provided, using solar metallicity
and included the following 11 filters $B$, $V$, $i'$, $z'$, $Y$, $J$,
$K_{s}$, and the four IRAC channels . We employed option 2 for the error
treatment of an undetected source which assumes that the flux in that
filter and its $1\sigma$ error are equal to half the flux of the
limiting magnitude (i.e. the error bar ranges from flux=0 to the
$1\sigma$ limiting magnitude in that waveband). 

\noindent{\bf Object 2200}

This object displays a strong detection in the $Y-$, $J-$ and $K_{s}-$bands
as well as a significant detection in the IRAC channels 1 and 2. There
is some flux detected in the $z'$-band, however this is to be
expected for some candidates as an examination of Figure
\ref{fig:filters} shows the $z'-$ and $Y-$filter transmission curves do
overlap significantly. This means as the Lyman break moves through the
$Y-$band filter with increasing redshift its contribution to the
$z'$-band flux will decrease but may not entirely disappear.  The $K_{s}-$band source is slightly offset from the detection in the other wavebands for object
2200, this prompted us to widen our search area in the GOODS-MUSIC
catalogue to a $1\farcs0$ radius. This larger radius yielded a match to
an object in GOODS-MUSIC with an ID 30199 and $z_{phot}=2.73$. This object was also
identified by Stanway et al. (2008) as a possible $z'$-drop. However,
the detection by Stanway et al. is centered $0\farcs8$ from our source.

This object also has a reported MIPS $24\,\mu$m detection
(source mip003485 in Alonso-Herrero et al.\ 2006), and
coincides with a $0.5-2$\,keV {\em Chandra} X-ray source,
and hence probably has an AGN contribution.
This object has also been presented in Dunlop, Cirasuolo \& McLure (2006),
their object 2336, who derive a photometric redshift in
the range $z=3.25-3.85$.  The photometry presented in Table~\ref{tab:zcandidates} is for an object at the position of the $Y-$band source, this is offset from the $K_{s}-$band source and as such has a fainter magnitude than that given by Dunlop et al. (2006). We believe that the $Y-$band detection is probably associated with the $K-$source, although this requires additional imaging and/or spectroscopy to confirm.


\noindent{\bf Object 9136}

\begin{figure}
\resizebox{0.48\textwidth}{!}{\includegraphics{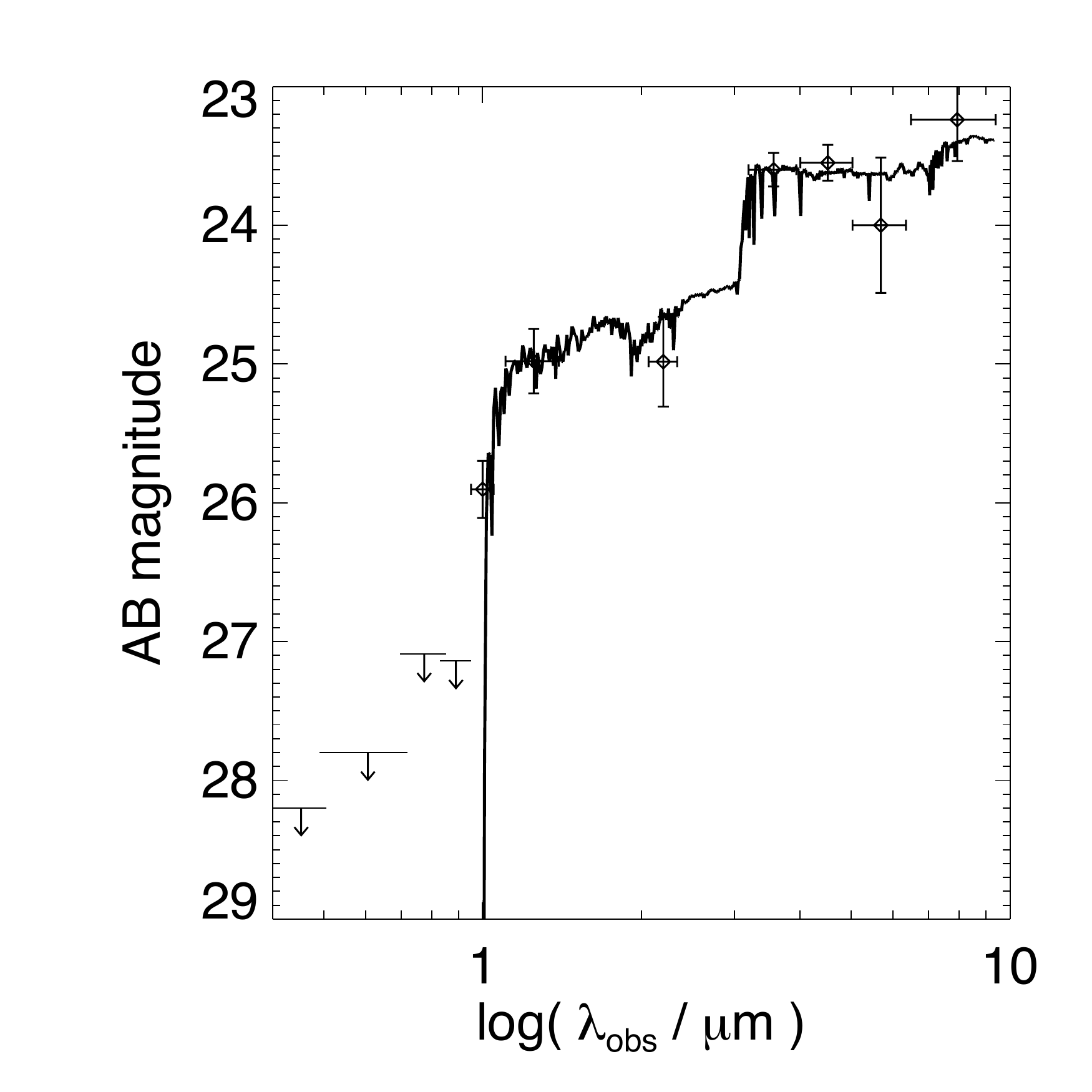}}
\caption{Best fit SED at $z_{phot}=7.01$ to Object 9136 with photometry overlaid and 2 sigma upper limits denoted by down arrows}
\label{fig:sed9136}
\end{figure}

\begin{figure}
\resizebox{0.48\textwidth}{!}{\includegraphics{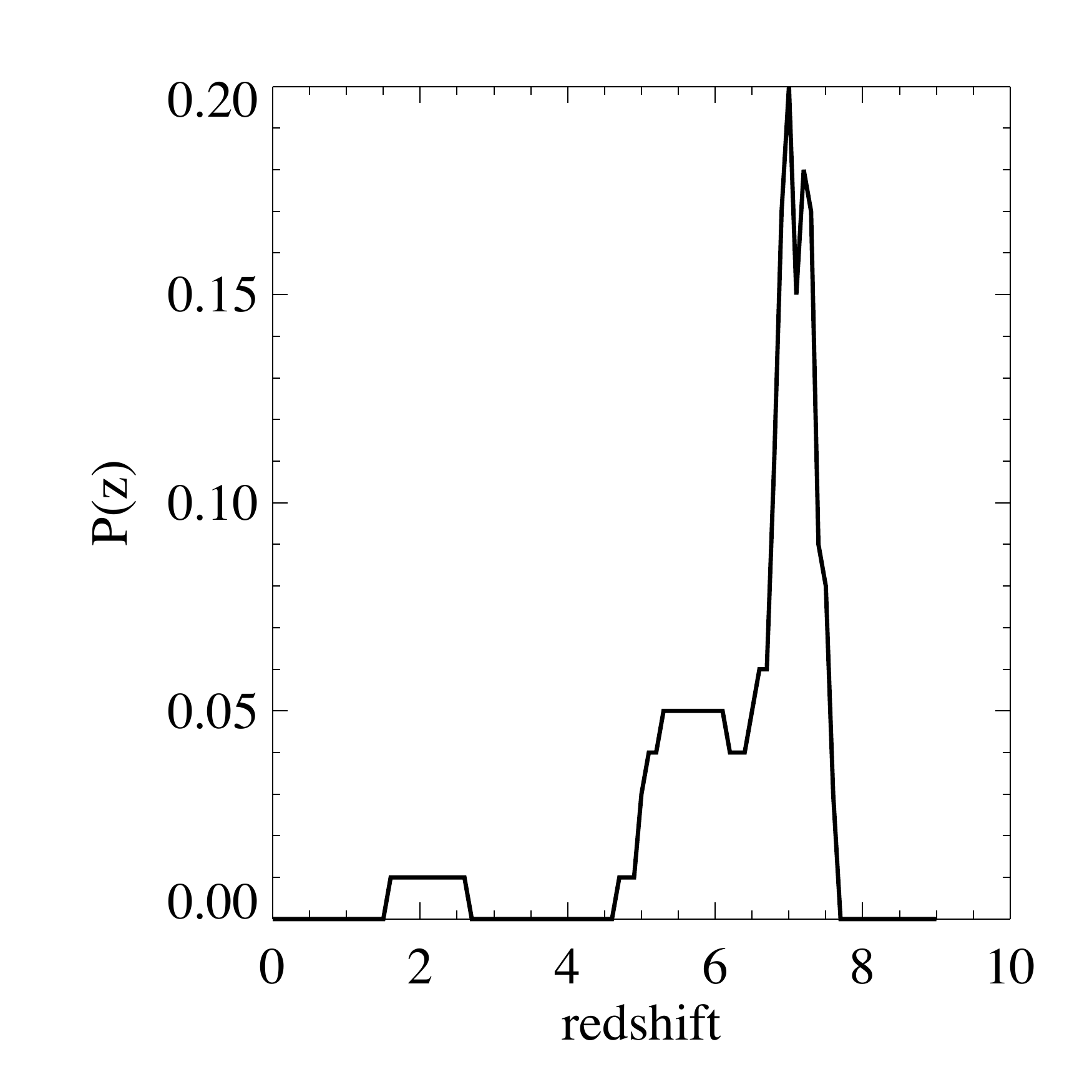}}
\caption{The redshift probability distribution for Object 9136, showing the best-fit photometric redshift of $z=7.01$.}
\label{fig:prob9136}
\end{figure}

This candidate displays a strong detection in the $Y$ band and is also
detected in the $J$ and $K_{s}$bands, and like object 2200 it is strongly
detected at the IRAC wavelengths.  In Figure \ref{fig:sed9136} we
show the photometric data points for this object with the best fit galaxy
template overlaid. The best fit solution from {\sc Hyperz} is $z_{phot}=7.01$ with a
secondary peak in the probability distribution at $z_{phot}=7.23$ 
(Fig.~\ref{fig:prob9136}). This best fit solution is for a starburst
galaxy with $A_V=0$.  We calculated star formation rates for our
objects based on the rest frame UV continuum at $1500$\,\AA\ (see
Bunker et al. 2004). For object 9136 we find a star formation rate of
$\approx 22$ M$_{\odot}\,{\rm yr}^{-1}$.

\noindent{\bf Object 9266}

This object is detected in the $Y$-band, but not in the $J$- and $K_{s}$-bands. However the limits in these bands are fainter than the measured $Y$-band magnitude. This could indicate that the object is spurious or a result of line contamination in the $Y-$band filter, or simply that its continuum is fainter than the $J-$ and $K_{s}-$band limits but bright enough to be detected in the $Y$-band.
We did not fit a photometric redshift for this object due to its limited detections.

\noindent{\bf Object 9697}

\begin{figure}
\resizebox{0.48\textwidth}{!}{\includegraphics{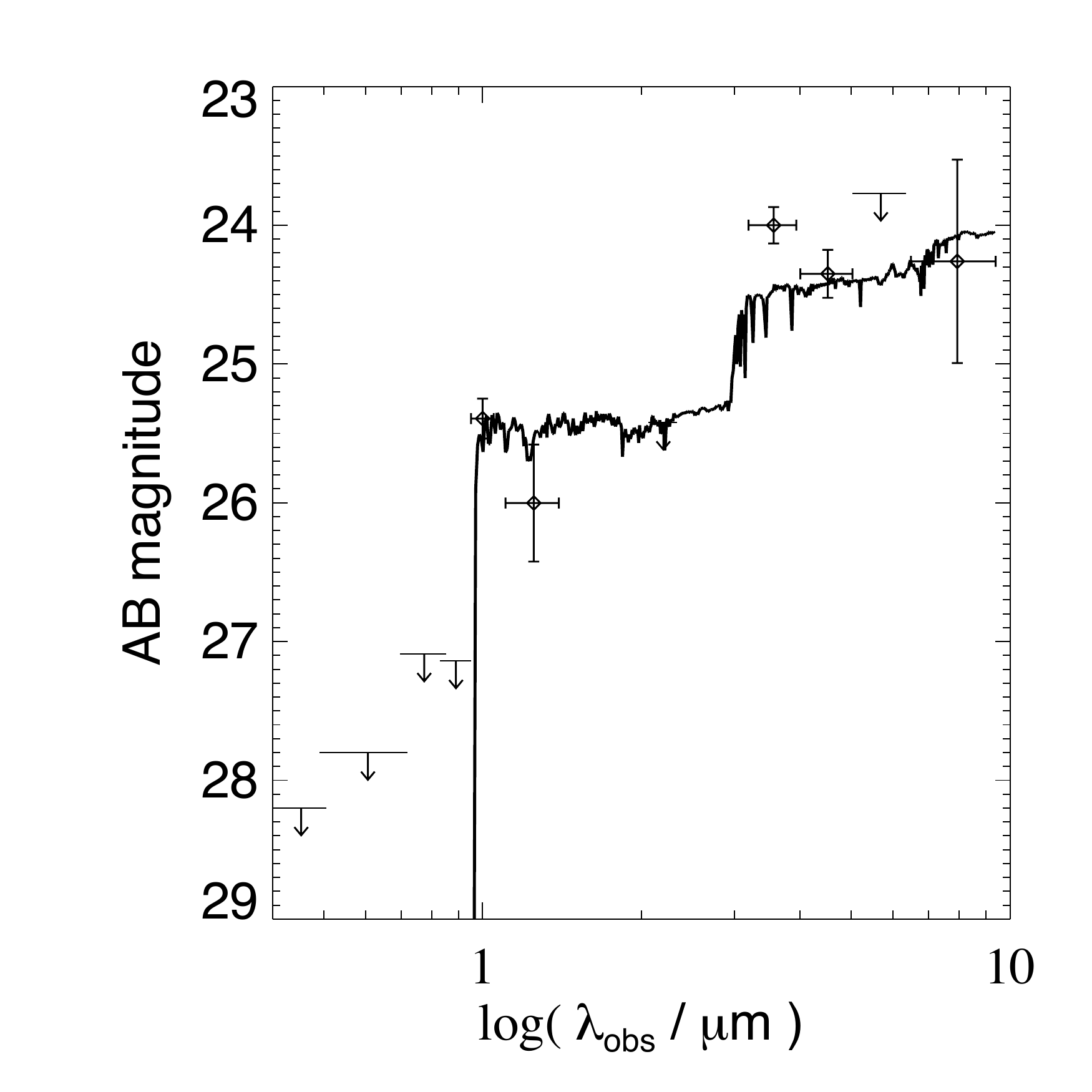}}
\caption{The best-fit SED at $z_{phot}=6.92$ for Object 9697 with the multiband photometry overlaid. The $2 \sigma$ upper limits are denoted by the arrows}
\label{fig:sed9697}
\end{figure}

\begin{figure}
\resizebox{0.48\textwidth}{!}{\includegraphics{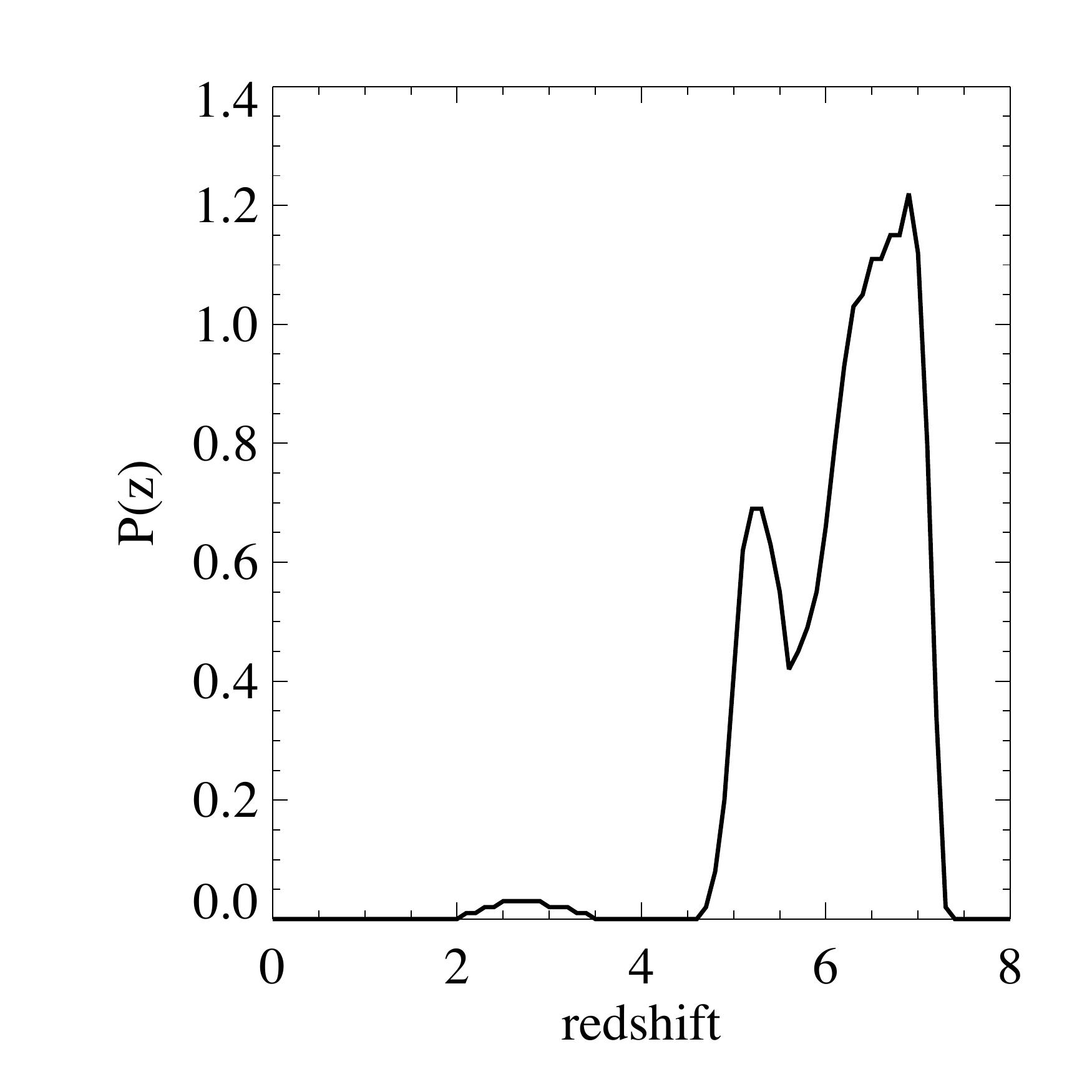}}
\caption{The redshift probability distribution for Object 9697, which shows the best-fit photometric redshift of $z=6.92$.}
\label{fig:prob9697}
\end{figure}

This object was previously identified as an $i'$-band drop-out in the
GOODS ACSv1 data by Bouwens et al.\ (2006), and is \#2226643007 in
their catalogue, with $z'_{AB}=27.54 \pm 0.18$, $(i'-z')>1.3$ and infrared
magnitudes $J_{AB}=26.04$ and $K_{AB}>25.4$.  The source is detected
in the $Y$- and $J$-bands and there is also a strong {\em Spitzer}
detection. However there is a nearby source in the $z'$-band
unconfirmed at the other wavelengths and another detection $\approx
0\farcs7$ to the east in the $B$ and $V$ bands which, while
unassociated with the $z'$-band detection, may be at least partially
responsible for the IRAC flux as it does fall within the IRAC
aperture. However it would be a rather unusual object to be detected
in $B$ and $V$ and the {\em Spitzer} bands and undetected in $z'$,
$Y$, $J$, and $K_{s}$.  The $i'z'YJK_{s}$ colours appear consistent with a
high-redshift interpretation.  We also ran {\sc Hyperz} on this object
 assuming the {\em Spitzer} flux was contributed by our candidate to
determine its photometric redshift. In Figure \ref{fig:sed9697} we
show the photometry of this object with the best-fit galaxy template
overlaid. The best-fit solution was $z_{phot}=6.92$ with a secondary
peak in the probability distribution at $z_{phot}=5.22$, see Figure
\ref{fig:prob9697}. This best-fit solution is for an elliptical galaxy
with $A_V=0$. We calculate the star formation rate based on the UV
continuum to be $\approx 25 M_{\odot}\,{\rm yr}^{-1}$.

Another possibility, other than a high redshift interpretation, is that the objects which are strongly detected in the
IRAC bands, 9136, 9697 and 2200 may be similar to IRAC-selected
extremely red objects (IEROs) see (Yan et al. 2004). But in the sample
discussed by Yan et al. the sources had optical detections which our
candidates do not. At the high-redshifts that we estimate for these galaxies, if there was an old
stellar population present, the 4000\,\AA\ break could fall between
the $K_{s}$ and the $3.6 \mu$m bands. This could explain some of our
sources increasing in brightness at the IRAC wavelengths.
 
\begin{figure}
\resizebox{0.48\textwidth}{!}{\includegraphics{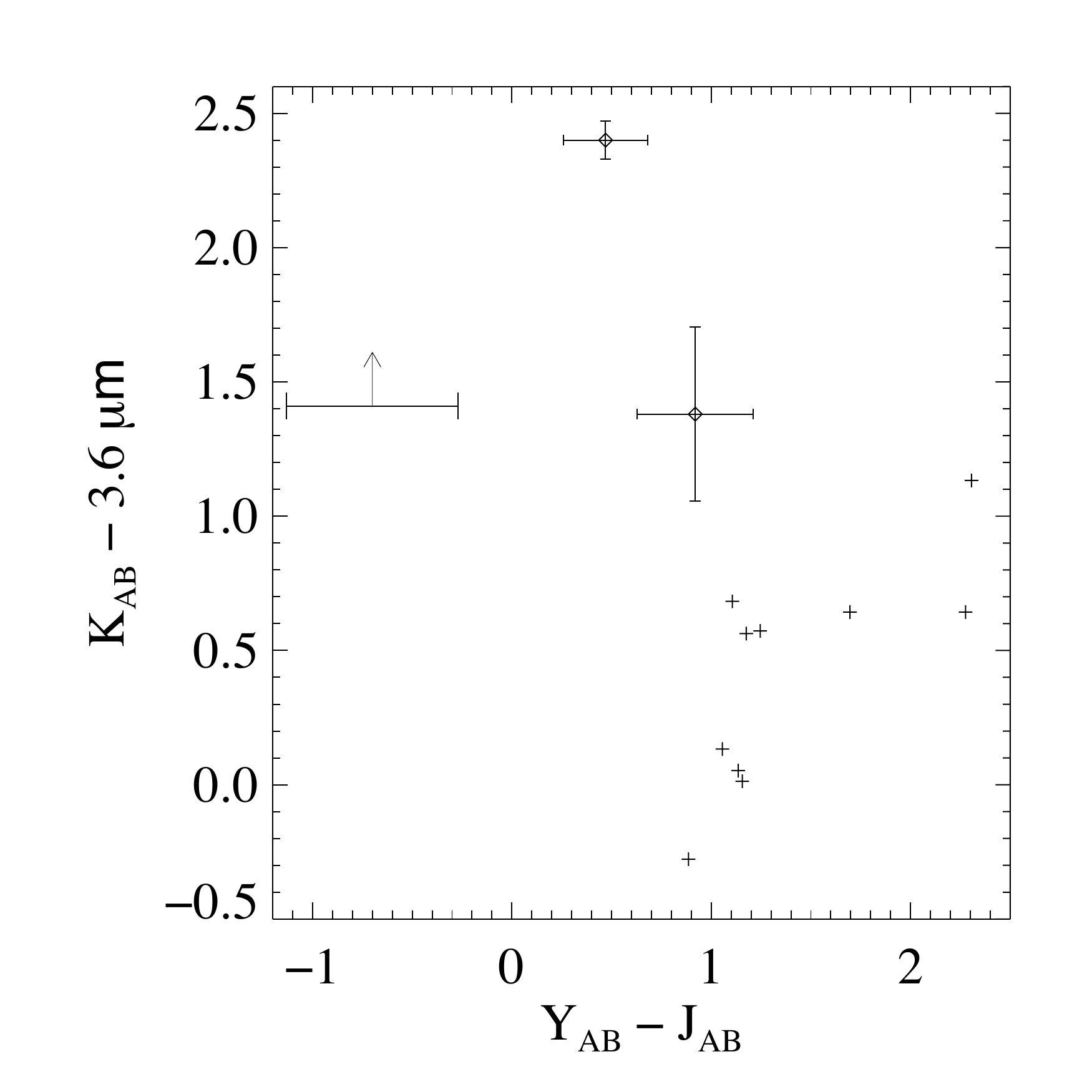}}
\caption{$Y-J$ vs. $K- 3.6$ colours for a sample of brown dwarfs from L1-T8. The points on the plot mark our three of our candidates, from left to right 9697, 2200 and 9136. Object 9266 is not shown as it is only detected in the $Y$ band. All 3 of these candidates lie away from the brown dwarfs marked by plus signs. Brown dwarf colours are taken from Leggett et al. (2000, 2001,2002), Kirkpatrick et al. (2000), Geballe et al. (2002), Knapp et al. (2004) and Hewett et al. (2006).}
\label{fig:browndwarf}
\end{figure}

We also explored the possibility that our remaining candidates could
be brown dwarfs. Patten et al. (2006) presented observationally-derived colours for various spectral types of M, L
and T dwarfs in the near-infrared and IRAC bands. In order to rule out our
candidates as brown dwarfs we compared their near-infrared colours to the
following three colour spaces: $[3.6-4.5]$; $[J-4.5]$; and $[K_{s}-4.5]$. We
combined the errors on our data points to conservatively explore the
colour space covered by our objects and compared them with the
expected brown dwarf colours. We found $[3.6-4.5]$ colours indicative
of spectral types T2-5 for object 9136 but we found $[J-4.5]$ colours
indicative of an L8 dwarf and $[K_{s}-4.5]$ colours of a T7 or T8
dwarf. Each constraint contradicts the next, therefore we find it
unlikely that object 9136 is a brown dwarf.

Object 9697 has $[3.6-4.5]$ colours consistent with all M, L and T0-3
spectral types  and $[Ks-4.5]$ colours consistent with L5-8 and all
T-dwarfs and (at the limit of its errors) $[J-4.5]$ colours between $3.47$ and $4.59$. The plots provided be Patten et al. (2006) indicate that a brown dwarf of type M, L or T will have $[J-4.5]$ colours of $< 3.5$, so at the extreme of its errors, object 9697 has colours just
consistent with an L8 brown dwarf. However when we compared typical $Y-J$  colours of brown dwarfs (Hewett et al. 2006) with $K_{s}-3.6 \mu$m colours, see Figure \ref{fig:browndwarf}, we find that object 9136 lies significantly away from the typical low-mass star colours, again making it unlikely that our objects can  be explained as brown dwarfs.
 A combination of resolution (in the near-infrared bands) and signal-to-noise ratio (in the $z'-$band) is insufficient to determine whether this object is unresolved, as would be expected for a brown dwarf.

\subsection{Plausibility of $Y$-drop candidates}

Our final list of $Y$-drop candidates consists of 4 objects. They span
a magnitude range of $J_{AB}=25.0-25.4$ after applying aperture
corrections. 

\noindent {\bf Object 2058} is detected only in the $J$-band with no IRAC source associated.

\noindent{\bf Object 4551} is also only detected in the $J$-band but falls in a noisy region of the $Y$-band image.

\noindent {\bf Object 5512} falls on the edge of the ACS images. 

\noindent{\bf Object 4532} has a $J$-band detection which is possibly
associated with the extended edge of a galaxy $1\farcs2$ away. This
source is detected in all of the ACS bands and has a match to
GOODS-MUSIC object 9610 with a spectroscopic redshift of $z=0.347$. It
is likely that the large apparent $(Y-J)$ colour recorded at the
position of object 4532 is due to worse seeing in the $J$-band than
the $Y$-band, or perhaps an intrinsic colour gradient in
GOODS-MUSIC\,9610. Given the proximity of this low-redshift source, it
is extremely unlikely that object 4532 is a genuine $Y$-drop at $z>8$.

We find no robust $Y$-drop candidates as all of our candidates only
appear in the $J$-band image with no significant $K_{s}$-band detection
which is unexpected as the $J$ and $K_{s}$-bands probe similar
depths. They also have no clear IRAC detection. This could indicate
spurious detections in the $J$-band because they are unconfirmed in
any other, or it could be the result of line contamination in the
$J$-band filter from\ high equivalent width Lyman $\alpha$. They
could also be galaxies with very blue spectra indicating low
metallicity and very little or no dust.


In order to assess whether such blue colours could be plausible, we
consider two scenarios, one of which has the colour difference
produced by a blue spectral slope, the second assumes that the
brightness in the $J$-band relative to $K_S$ is attributable to a very
strong emission line (e.g. Lyman-$\alpha$ at $8.0<z<10.5$).  We then
compare the constraints on the spectral slope and Lyman-$\alpha$
equivalent width with the known properties of Lyman-break galaxies at
high redshift. If our limits fall outside the range observed in
distant galaxies, then the $J$-band detections are probably
spurious (i.e., inconsistent with these being $Y$-band drop-outs at
$z>8$).

We assume a simple power law for the spectrum of $f_{\lambda}\propto
\lambda^{\beta}$ (or equivalently $f_{\nu}\propto
\lambda^{\beta+2}$, where $\beta=-2$ is a spectrum of constant $AB$
magnitude, flat in $f_{\nu}$). We place $2\,\sigma$ upper limits on the
spectral slope of $\beta<-2.44$ (for sources 4551 and 5512 with
$J_{AB}=25.0$) and $\beta<-2.06$ (for sources 2058 and 4532 with
$J_{AB}=25.4$); these limits are conservative, because if the
Lyman-$\alpha$ break occurs in the $J$-band filter (i.e. $z>8$) rather
than shortward of $1.1\,\mu$m then the true spectral slope would be
even bluer.  Our $2\,\sigma$ limits are consistent within the errors with the reported
average for $z=6$ $i'$-drop galaxies. Stanway, McMahon \& Bunker
(2005) derive values of $\beta=-2.2 \pm 0.2$ from $i'$-drop galaxies at
$z\approx 6$, with Bouwens et al.\ (2008) reporting $\beta =
-2.0$. 

As an alternative to the blue $(J-K_S)$ colour being due to a steep
blue spectral slope, we now consider whether the apparent flux excess
in the $J$-band could be due to emission line contamination.  To
determine lower limits on the equivalent width of this putative line
emission, we assume a spectrum flat in $f_{\nu}$ (i.e., constant AB
magnitude with wavelength) longward of Lyman-$\alpha$; this is typical
of a low-extinction star-forming galaxy. We take our $2\,\sigma$ upper
limit for the flux density ($f_{\nu}$) in the $K_S$-band as the upper limit
on the continuum level in the $J-$band filter, and attribute the
$>1.4\times$ greater flux density (for objects 4551 and 5512) in the $J$-band ($2\,\sigma$
lower limit) to be due to line emission. The
$J$ band has a width of 3000\,\AA , which sets a $2\,\sigma$ lower
limit on the observed equivalent width of $EW_{obs}>1205$\,\AA\ (for
$J$-drops 4551 \& 5512). This corresponds to a rest-frame equivalent width of $EW_{rest}>133$\,\AA\ if the line is Lyman-$\alpha$ at $z=8$, where it enters the $J$-band (this limit is
conservative because at larger redshifts, an increasing fraction of
the $J$-band falls below the Lyman-$\alpha$ break, so the contribution
of the continuum to the flux density would be even lower and hence the
equivalent width higher). These equivalent widths are plausible for the Lyman-break and Lyman-$\alpha$ emitter populations at high redshift (e.g., Dawson et al. 2004). We can rule out
lower redshift lines with higher confidence (e.g., H$\alpha$ at
$z\approx 0.9$ would have an implausibly high
$EW_{rest}>634$\,\AA , and [O{\scriptsize~II}]\,3727\,\AA\ at
$z\approx 2.3$ would have $EW_{rest}>365$\,\AA ).

We regard the marginal $J$-band detections to be highly suspect, 
although we cannot eliminate them from our selection based on their blue spectral slopes or high equivalent width line emission falling in the $J$-band. They are unconfirmed in any other band studied here and may be the result of spurious detections.  Deeper imaging in $J$, $H$ and $K$ and/or spectroscopy is required to confirm or disprove the nature of these candidates.

\begin{figure*}
{\includegraphics{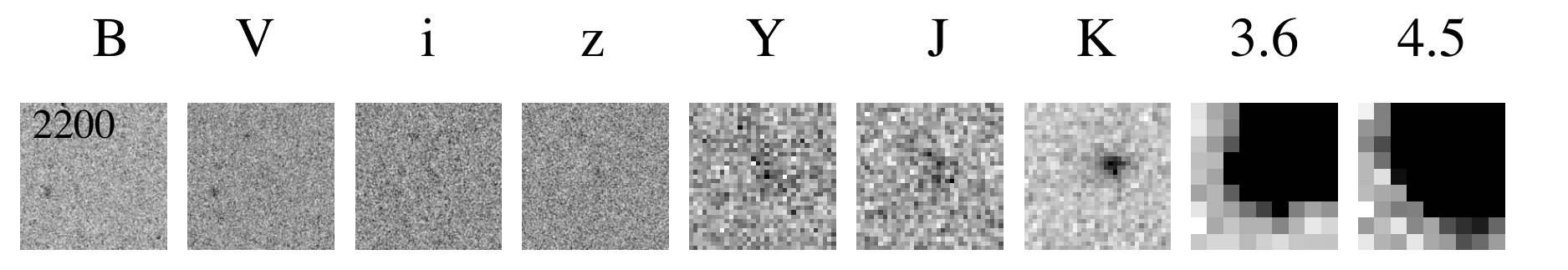}}
{\includegraphics{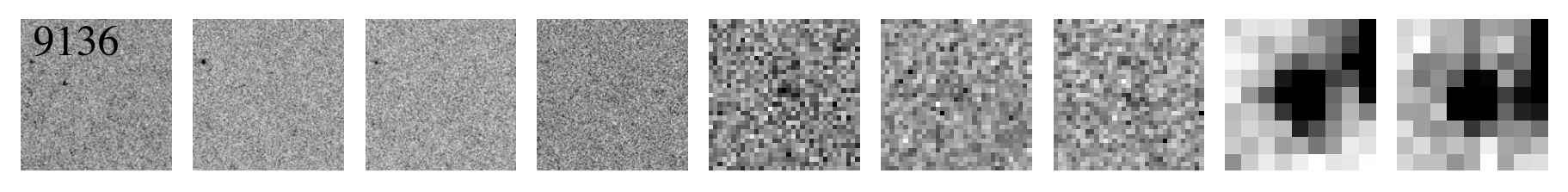}}
{\includegraphics{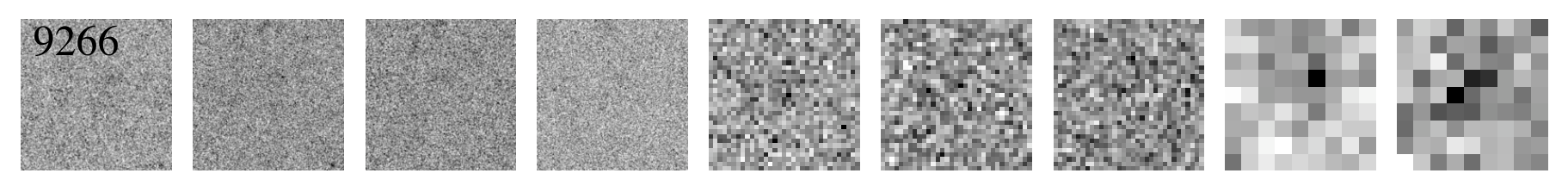}}
{\includegraphics{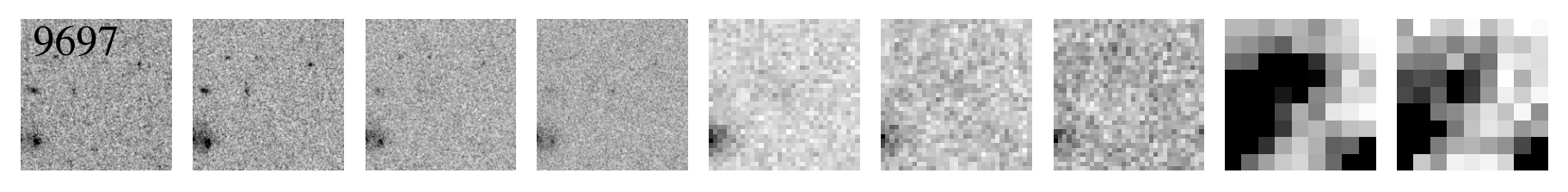}}
\label{fig:zdrops}
\caption{$z'$-band drop-out candidates. Each postage stamp is 5x5 arcsec in size. All candidates are undetected in the optical wavebands at the $2 \sigma$ level but have $>5 \sigma$ detections in the $Y$-band. Objects 9136 and 9697 are detected in all of the longer wavelength filters,  excluding the $Ks$-band for Object 9697, making these our most plaussible candidates. }

\end{figure*}

\begin{figure*}
{\includegraphics{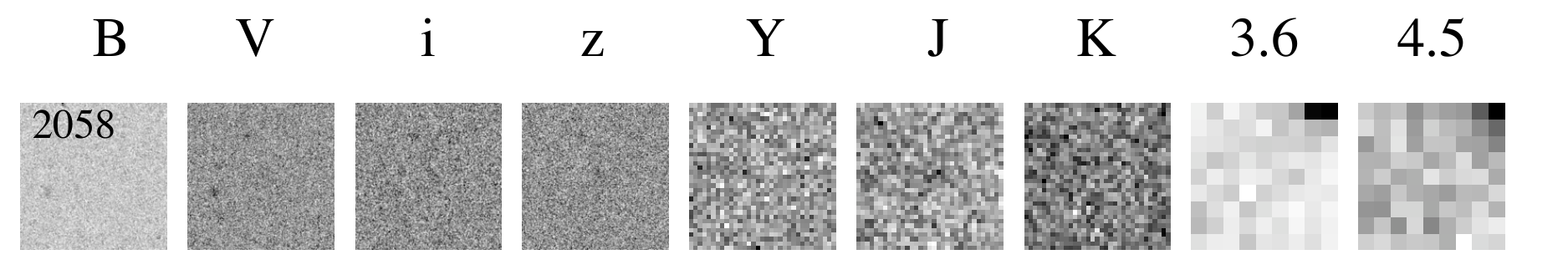}}
{\includegraphics{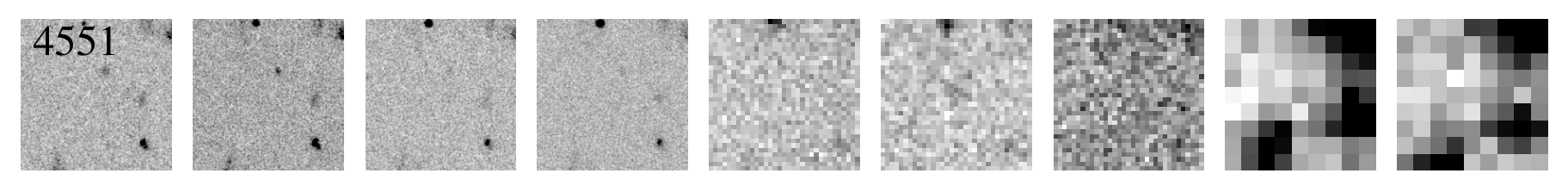}}
{\includegraphics{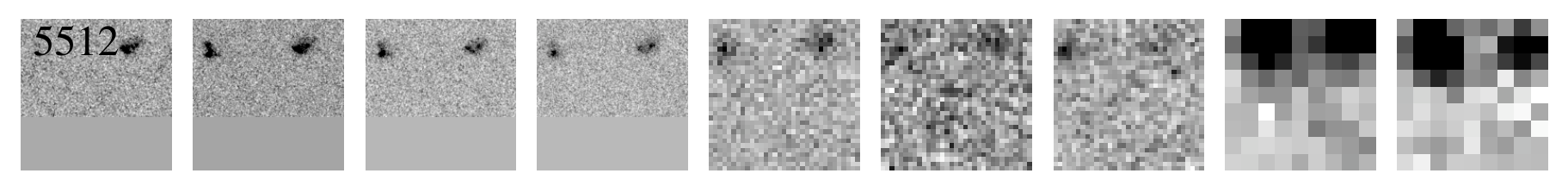}}
{\includegraphics{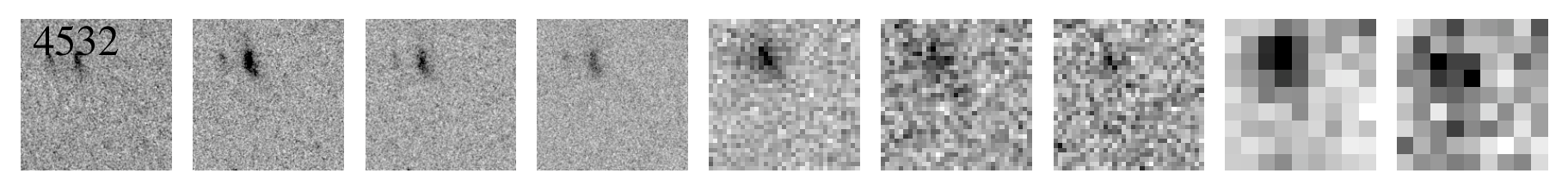}}
\label{fig:ydrops}
\caption{$Y$-band drop out candidates. Each postage stamp is 5x5 arcsec in size.  The candidates are undetected in the optical wavebands down to the $2 \sigma$ level. Each $Y$-drop candidate is detected in the $J$-band with $S/N > 5$ but have no clear detection in any of the longer wavelength bands.}

\end{figure*}

\subsection{Discussion of other work}
\label{sec:otherwork}
There are a few authors who have provided galaxy candidates at $z>6.5$
using the Lyman Break technique. As mentioned before, one of our
$z'$-drop candidates, object 2200 has also been identified by Stanway
et al. (2008) but with a slight offset of $0\farcs8$. Two other
objects were also identified as possible high-redshift candidates in
that paper, one of which we would not expect to detect because it does
not lie within our field and the other object has a signal-to-noise ratio 
of less than our cut of $S/N > 5$ in the $Y$-band. Within the
limits of their study, Stanway et al. found the luminosity function at
$z=7$ to be consistent with predictions of the luminosity function
from Bouwens et al.\ (2007) and McLure et al.\ (2009) at $z\sim 6$.  Two other objects within the
GOODS-South field were originally flagged by Mannucci et al. (2007) as
high-redshift candidates but were later dismissed as brown dwarfs
based on their morphologies, {\em Spitzer} colours and spectroscopic
information. These objects were included in our original catalogues as
they do have the colours of high-redshift galaxies, and they were also
identified in the GOODS-MUSIC catalogue with photometric redshifts of
$z\approx6.9$ (objects 11002 \& 7004 in the GOODS-MUSIC
catalogue). Based on the non-detection of any credible candidates
Mannucci et al.\ placed constraints on the UV luminosity function at
$z=7$ and claimed strong evolution in the luminosity function from
$z=6$ to $z=7$.  Further evidence for the evolution of the luminosity
function from $z=3.8$ to $z=6 \rightarrow 7$ is presented in Bouwens et al. (2004,
2005, 2008). No robust $J$-drops were presented but a number of
$z'$-drop candidates were found. However the observations used in the
Bouwens et al. (2008) study were much deeper than in other searches
and some results even implied evolution from $z=6$ to $z=7$ and a
potential luminosity function at $z=7$ was derived while constraints
were set on the luminosity function at $z=9$ (the $J$-drop
population).

\subsection{Implications for the UV Luminosity Function}

The number of robust candidates we detect can constrain the UV
luminosity functions at $z> 7$.  
We compute the number of galaxies expected to be selected within our
survey area for several luminosity functions, 
derived from lower-redshift samples. A significant discrepancy
between our observed number counts and those predicted would
argue for strong evolution in the star-forming population with
redshift.

To model the predicted number counts, we first adopt a simple model
spectrum of a star-forming galaxy, where the rest-UV spectrum is
approximately flat in $f_{\nu}$ longward of Lyman-$\alpha$ (i.e.,
$\beta=-2$ where $f_{\lambda}\propto \lambda^{\beta}$, appropriate for
star-forming galaxies at $z\approx 6$ -- Stanway, McMahon \& Bunker
2005) and is severely attenuated below Lyman-$\alpha$ due to the
opacity of the intervening neutral hydrogen absorbers (we adopt an
absorption of $D_A=0.99$ for $z>6.5$). The $Y$-band filter is sensitive to the UV continuum longward of Lyman-$\alpha$ at
$6.6<z<7.7$, although at the higher redshifts the galaxies would have
to be extremely luminous to appear in our magnitude-limited
sample -- not only is the luminosity distance greater, but also a
smaller fraction of the filter bandpass lies above Lyman-$\alpha$. We
model this by considering small increments of redshift ($\Delta
z=0.1$) between $z=6$ and $z=8$, and for each redshift bin we calculate
the number of galaxies expected as a function of limiting
apparent magnitude. We also determine the expected $z'-Y$ colours to
assess whether our colour cut will select star-forming galaxies in
that redshift bin.  By summing over all the redshift bins, we obtain
the expected surface density of $z'$-drop galaxies as a function of
magnitude. We then consider our exposure maps, and use this to compute
the various areas of sky observed to different limiting
magnitudes. For each area
observed, we correct the predicted number counts for our measured
completeness as a function of magnitude (see Section
\ref{sec:comp}). By summing the expected number of galaxies above
our colour cut ($z'-Y >1.0$) and our significance threshold ($S/N>5$ in
$Y$-band) for each area of the survey, we obtained the total number of
$z'$-drop star-forming galaxies we would expect to find if the assumed
luminosity function is appropriate at $z\approx 7$.

We can compare the number of expected galaxies derived from models  
with the number we actually detect.  In this paper we will compare two
UV luminosity functions, one derived by Steidel et al (1999) for
Lyman-break galaxies at $z=3$ (the $U$-band dropouts), and the other
by Bouwens et al.\ (2007) for the Lyman-break population at $z\approx
6$ (the $i'$-band dropouts). The Steidel et al.\ UV luminosity
function at $z=3$ has a faint end slope of $\alpha=-1.6$ and
$L^{*} _{SFR}=15.0\,M_\odot$\,yr$^{-1}$ and $
\Phi=0.00138$\,Mpc$^{-3}$, where $L^{*}$ is derived from the
rest-frame UV around $1500$\,\AA\ and has been converted to an
effective star formation rate using the relation $L_{UV}=8\times
10^{27}\times SFR{\rm ~\,ergs\,s^{-1}\,Hz^{-1}}$ (Madau, Pozzetti \&
Dickinson 1998), appropriate for a Salpeter (1955) stellar initial
mass function.  At $z\approx 6$ the Bouwens et al.\ luminosity
function shows strong evolution in $L^{*}$ from $z\sim3$, with $L^{*}
_{SFR}=8.6\,M_\odot$\,yr$^{-1}$ (equivalent to $0.575\,L^{*}_{UV}$ at
$z=3$). The faint end slope is also steeper at $z\approx 6$
($\alpha=-1.74$) and $\Phi=0.001135$\,Mpc$^{-3}$ (which is
$0.82\,\phi^{*}_{z=3}$).

\begin{figure}
\resizebox{0.48\textwidth}{!}{\includegraphics{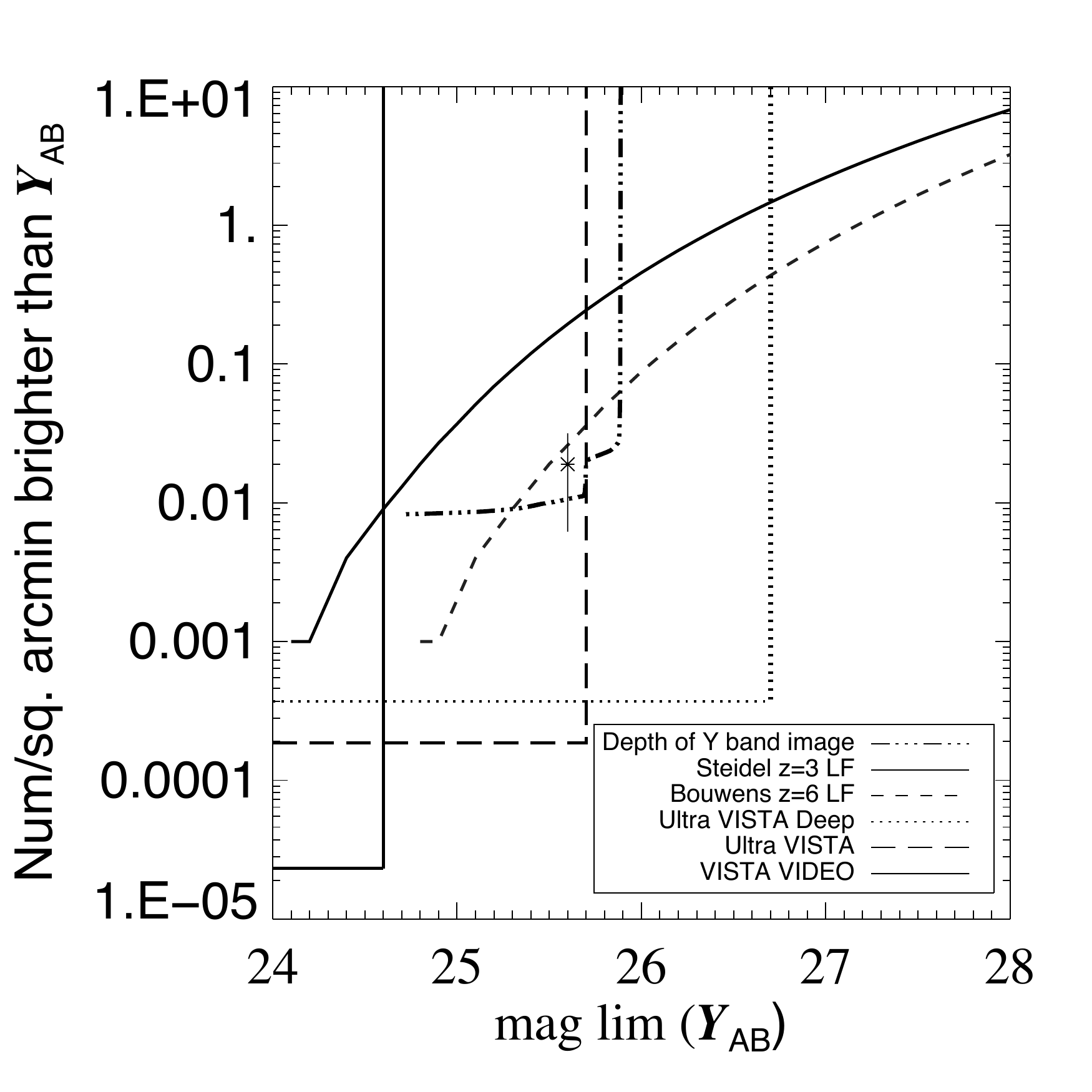}}
\caption{Expected number of $z'$-drops, with the solid line assuming the $z=3$ luminosity function from Steidel et al.\ (1996) and the dashed line the $z=6$ luminosity function from Bouwens et al.\ (2006). The dotted dashed line marks the phase space probed by the $Y$-band data. The point denotes the two candidates we found. The dotted, hyphenated and solid lines mark the phase spaces that will be probed by the Ultra VISTA Deep, the Ultra VISTA shallow and the VISTA VIDEO surveys respectively. The luminosity functions shown here are not constrained at the bright end but with the depth and area of the VISTA surveys we will be able to measure the form of the function at these bright magnitudes.}
\label{fig:numdens_zdrop}
\end{figure}

For the deepest region of our survey (pointing 1), we expect a
$z'$-drop surface density brighter than our 50 per cent completeness
limit ($Y_{AB}<25.9$) of $0.373\,{\rm arcmin}^{-2}$ and $0.066\,{\rm
  arcmin}^{-2}$ for the Steidel et al.\ (1999) and Bouwens et al.\
(2007) luminosity functions respectively.  Accounting for completeness
and the different depths as a function of survey area, the total
numbers expected are $29.5\pm 5.4$ or $5.2\pm 2.3$ if the $z=3$ or
$z=6$ luminosity functions are appropriate for the Lyman-break
population at $z\approx 7$ (Figure~\ref{fig:numdens_zdrop}). Clearly, as we have only two robust
candidate $z'$-drops (and at most three), we can strongly rule out a
model where there is no evolution in the rest-frame UV luminosity
function from $z=7$ to $z=3$, as the number of high-redshift galaxies
is over-predicted by a factor of 10. There is some evidence for evolution from
$z=6$ to $z=7$: it is likely that some or all of out $z'$-drop
candidates are not at $z=7$, and hence the observed number is at least
a factor of two less than the prediction based on the $z=6$ luminosity
function, although the statistical significance of this is marginal
given the small numbers.

\begin{figure}
\resizebox{0.48\textwidth}{!}{\includegraphics{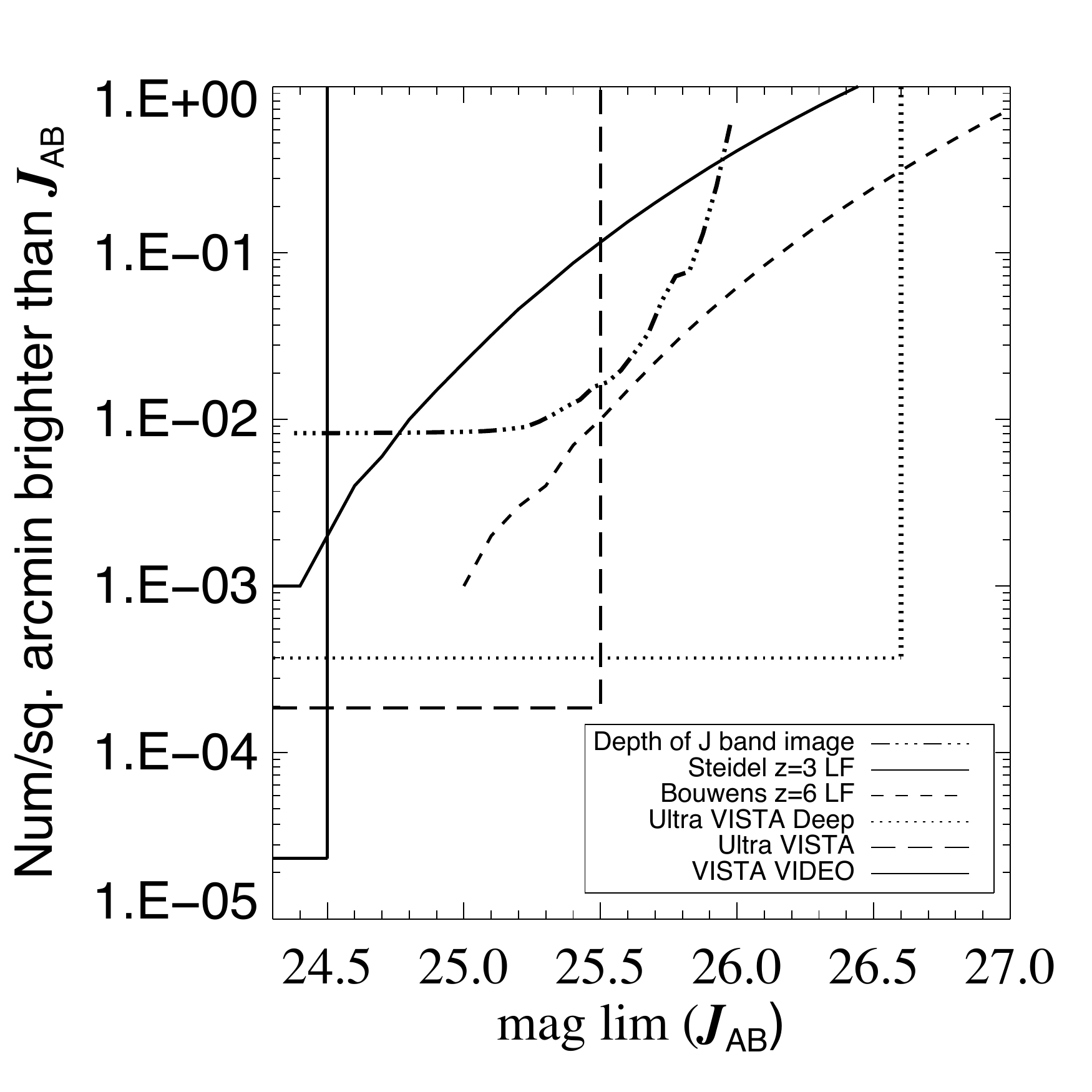}}
\caption{Expected number of $Y$-drops, the various lines are the same as those presented in figure~\ref{fig:numdens_zdrop} but using the $J$ band as the long wavelength detection band.}
\label{fig:numdens_ydrop}
\end{figure}

The same models were applied for the $Y$-drop candidates. Our 50 per
cent completeness limit in the $J$-band is at $J_{AB}=25.4$,
correcting for incompleteness means we expect to find $10.5\pm3.2$ and
$1.1\pm1$ $Y$-drops for the Steidel et al. and Bouwens et al. luminosity functions respectively
(Figure~\ref{fig:numdens_ydrop}).  All four of our candidates lie
very close to our $5\,\sigma$ $J$-band cut and object 4532 is possibly the
edge of an extended object $\approx 1\farcs2$ away. They are also
undetected in $K_{s}$-band which is of a similar depth to the $J$-band and
hence the sources may be spurious or the result of line contamination in the
$J$-band. Although we cannot rule out all of our $Y$-band candidates
we do not believe them with a high degree of confidence. Thus we are
again inconsistent with the Steidel et al.\ luminosity function
implying evolution in the UV luminosity function between $z=8$ and
$z=3$. Within the errors we are consistent with the Bouwens luminosity
function at $z=6$.

At the limit of our survey for the $z'$-drop at $z\approx 7$ we begin
to probe the $z=6$ Bouwens et al. luminosity function if there is little evolution, but in order to
constrain the luminosity function more effectively, deeper and/or wider
observations are needed. This is a possibility with the VISTA Deep Extragalactic Observations (VIDEO) and the Ultra-VISTA surveys (Arnaboldi et al. 2007; see also Figures \ref{fig:numdens_zdrop} and
\ref{fig:numdens_ydrop}). Due to its large area of $\approx 12$
sq. degrees, VIDEO will be able to probe the bright end of the
luminosity function and after five years will reach $5 \sigma$
limiting magnitudes of $Y_{AB}=24.6$ and $J_{AB}=24.5$. UltraVISTA goes considerably deeper (to $Y_{AB}=25.7$ and
$J_{AB}=25.5$ over 1.5 sq. degrees and $Y_{AB}=26.7$ and $J_{AB}=26.6$
over 0.75 sq. degrees) but over a smaller area than VIDEO, and
hence UltraVISTA will be more effective at measuring the position of the break and the slope of the faint
end of the luminosity function (see Figure \ref{fig:numdens_zdrop}).

\section{Conclusion}
\label{sec:concs}

We have searched for high-redshift drop-out galaxies in the GOODS-South field using the new HAWK-I $Y$-band data covering $\sim 119$~arcmin$^{2}$. We have complemented this data with VLT ISAAC $J$ and $K_{s}$ images in addition to HST-ACS images in $B$, $V$, $i'$ and $z'$ along with the deep {\em Spitzer} data in these fields. We employed a selection criteria of $(Y-J)_{AB}>0.75$ for the $Y$-drops and $(z'-Y)_{AB}>1.0$ for the $z'$-drops and a $S/N>5$. These catalogues were matched to the GOODS-MUSIC catalogue to eliminate objects with optical detections from our candidate lists.  Each remaining candidate was inspected by eye to eliminate remaining data artifacts, spurious sources and optical detections. A total of 4 $Y$-drop candidates were found within our data. Due to the fact that all of our $Y$-drop candidates are very close to our $S/N>5$ cut and are only significantly detected in the $J$-band, we do not believe any with a high degree of confidence. If none of our $Y$-band sources are indeed real, then this demands significant evolution in the UV luminosity function since $z=3$ based on the predictions by Steidel et al.(1999). 

We find 4 possible $z'$ drop candidates, one of which, Object 2200, has a probable low redshift solution of $z_{phot} = 3.25 - 3.85$. Another of our candidates, Object 9266, is only detected in the $Y$-band.  However we have 2 robust candidates, Objects 9136 and 9697,  which have significant detections in the IRAC wavebands and photometric redshifts of $z_{phot} > 6.9$. Our findings show evolution in the luminosity function since $z=3$ but are, within the Poisson error, consistent with the Bouwens et al. (2007) $z=6$ luminosity function although small number statistics preclude from making any strong statements on the evolution in the Lyman-Break population at $z> 6$. In order to  constrain the UV luminosity function at these high redshifts more effectively we need to search over a wider and/or deeper area, this will be possible with the combination of surveys such as VIDEO and Ultra VISTA (Arnaboldi et al. 2007).

\section*{Acknowledgments}

SH acknowledges a University of Hertfordshire studentship supporting this study and
MJJ acknowledges a Research Councils UK fellowship. 
We thank Ross McLure for useful discussions.
This
paper is based on observations made with the NASA/ESA Hubble Space
Telescope, obtained from the Data Archive at the Space Telescope Science
Institute, which is operated by the Association of Universities for
Research in Astronomy, Inc., under NASA contract NAS 5-26555. These
observations are associated with proposals \#9425\,\&\,9583 (the GOODS
public imaging survey). We are grateful to the GOODS team for making
their reduced images public -- a very useful resource. The archival
GOODS/EIS infrared images are based on observations collected at the
European Southern Observatory, Chile, as part of the ESO Large Programme
LP168.A-0485(A) (PI: C.\ Cesarsky), and ESO programmes 64.O-0643,
66.A-0572 and 68.A-0544 (PI: E.\ Giallongo).

\bsp

\label{lastpage}

\end{document}